\documentclass[final,3p,times]{elsarticle}
\usepackage{lineno,hyperref}
\usepackage{amssymb}
\usepackage{subfigure}
\usepackage{amsmath}
\usepackage{amsthm}
\usepackage{mathrsfs}
\usepackage{xcolor}
\usepackage{graphics}
\usepackage{amsmath,fancyhdr,amsthm,amssymb,amsfonts,version}
\usepackage{multirow}
\usepackage{bbm,version}
\usepackage{mathrsfs}
\usepackage{graphicx}
\usepackage{stmaryrd}
\usepackage{bm}
\usepackage{epstopdf}
\usepackage{float}
\usepackage{subfigure}
\usepackage{mathtools}
\usepackage{tabularx}
\usepackage{multirow,booktabs}

\modulolinenumbers[200]
\journal{Journal}

\begin{document}

\begin{frontmatter}

\title{Learning stiff chemical kinetics using extended deep neural operators}

\author{Somdatta Goswami$^{1,\dagger}$, Ameya D. Jagtap$^{1,\dagger}$, Hessam Babaee$^{2}$, Bryan T. Susi$^{3}$, and George Em Karniadakis$^{1,*}$}
\cortext[mycorrespondingauthor]{Corresponding author Emails: george$\_$karniadakis@brown.edu \\  $\dagger$ First two authors contributed equally.}

\address{$^1$ Division of Applied Mathematics, Brown University, 182 George Street, Providence, RI, 02912, USA

$^2$ Department of Mechanical Engineering and Materials Science, University of Pittsburgh, Pittsburgh, PA 15261, USA

$^3$ Applied Research Associates, Inc., Raleigh, NC, 27615-6545, USA}

\begin{abstract}
We utilize neural operators to learn the solution propagator for challenging systems of differential equations that are representative of stiff chemical kinetics. Specifically, we apply the deep operator network (DeepONet) along with its extensions, such as the autoencoder-based DeepONet and the newly proposed Partition-of-Unity (PoU-) DeepONet to study a range of examples, including the ROBERS problem with three species, the POLLU problem with 25 species, pure kinetics of a skeletal model for $CO/H_2$ burning of syngas, which contains 11 species and 21 reactions and finally, a temporally developing planar $CO/H_2$ jet flame (turbulent flame) using the same syngas mechanism. We have demonstrated the advantages of the proposed approach through these numerical examples. Specifically, to train the DeepONet for the syngas model, we solve the skeletal kinetic model for different initial conditions. In the first case, we parameterize the initial conditions based on equivalence ratios and initial temperature values. In the second case, we perform a direct numerical simulation of a two-dimensional temporally developing $CO/H_2$ jet flame. Then, we initialize the kinetic model by the thermochemical states visited by a subset of grid points at different time snapshots. Stiff chemical kinetics are computationally expensive to solve within the context of reactive computational fluid dynamics simulations, thus, this work aims to develop a neural operator-based surrogate model to efficiently solve stiff chemical kinetics. The operator, once trained offline, can accurately integrate the thermochemical state for arbitrarily large time advancements, leading to significant computational gains compared to stiff integration schemes.
\end{abstract}


\end{frontmatter}

\linenumbers

\section{Introduction}

The pace of discovery and advancement within the field of Scientific Machine Learning (SciML) is extraordinary, but equally impressive or even more so is the pace at which machine learning is successfully improving existing technologies in classical engineering applications. Across the many disciplines comprising machine learning research, applications have been explored in the various fields of computational science and engineering; see, for example, \cite{brunton2016discovering, wang2017physics,karpatne2017theory, chen2018neural,raissi2019physics}. This is significant because these diverse and independent computational modeling fields generally require different solution algorithms, different levels of numerical resolution, or even vary in their sensitivity to physical principles. The efficacy of machine learning for applications that need to solve systems of ordinary or partial differential equations (ODEs or PDEs) fast makes it particularly useful in the context of numerical modeling of transient combustion.

Computational Fluid Dynamics (CFD) \cite{ADITYA20192635} practitioners who model combustion processes recognize that the state of practice in their field is limited by contemporary computational power. There is a great disparity between the depth of collective knowledge about how to properly model transiently reacting fluid flows versus the accommodations and simplifications that are necessary given currently accessible computational resources. Be it the cost barrier for scalable cloud computing, the exclusivity of access to high-performance computing clusters, or even just finite memory or processing resources on personal computers, there are extant limitations that force various computational modeling fields to make compromises between what should be done from a physical perspective, and what contemporary computational resources can actually realize. This is because the physical processes for transient, non-premixed combustion lead to a multi-scale problem, spanning orders of magnitude differences in spatio-temporal scales. Subtle inaccuracies or under-resolution of the most minute Kolmogorov scales manifest as much larger inaccuracies in the predicted macro-scale fluid behavior. In a typical transient combustion simulation, computational models must resolve processes occurring on the order of nanoseconds and at sub-micron spatial scales to properly capture relevant macro-scale effects that occur over milliseconds or more, and at meter-length scales. CFD codes can resolve these phenomena, but either the computational resources required or the wall-clock time to wait for a simulation preclude simulations of anything but simplified scenarios at such high resolutions. This severely limits the computational fidelity for realistically sized scenarios.

Many techniques to alleviate the limitations imposed by stiff chemical reaction systems have been explored within the combustion modeling field. Time step splitting is still common practice to separate the chemistry evolution from the hydrodynamic evolution for example. It is also common to reduce full chemical kinetics models to their most skeletal form for more performant computations \cite{nouri2022skeletal} or to further reduce certain reaction system equations through Quasi-Steady State (QSS) assumptions \cite{goussis2012quasi, mott2000quasi} to help alleviate system stiffness. There are also procedures to bypass chemistry integration completely if possible, using techniques like Pope’s In-Situ Adaptive Tabulation (ISAT) scheme \cite{pope1997computationally, lu2009improved}. What these approaches all share in common, is an objective to reduce the computational burden of integrating finite rate chemistry. They vary in how this is achieved and in the consequences of their assumptions, but together they demonstrate the prevalence of simplifying combustion modeling problems at the cost of physical fidelity. The skeletal model and QSS approaches minimize the effects of deviating from a full mechanism, but fundamentally, they change the kinetic model. This could affect accuracy, but it also begets uncertainty in the simulation because the effect of the kinetic system simplification becomes one more variable that needs to be quantified to validate the simulation's results. Tabulation schemes like ISAT allow for error control such that inaccuracy can theoretically be minimized, but in the limit of zero introduced error, these methods are actually slower than finite rate chemistry because not much of the chemistry is being skipped but additionally the lookup step becomes more expensive as the state space grows. 
New approaches that enable very fast computation of changes to thermochemical states without compromising accuracy are highly desired, and Scientific Machine Learning is well suited to significantly improve the size and scope, or the speed, of transient reactive CFD simulations for combustion modeling. 

Holistic approaches of using Physics Informed Neural Networks (PINNs) \cite{raissi2019physics}, Deep Operator Networks (DeepONets) \cite{lu2021learning}, or other deep learning approaches for emulating the solutions of ODEs or PDEs in their entirety have yielded impressive results. Methods from SciML offer many opportunities to propel combustion modeling research toward greater physical fidelity, but completely dispensing with conventional continuum PDE solvers like spectral methods or finite element/volume approaches is unlikely at this juncture. Reactive CFD is one application where the variety of specific problem parameters and variations in computational topologies span an intractable training set from which to learn entire solutions. Formulating SciML methods specifically for finite rate chemistry, however, is advantageous for multiple reasons. The first and most notable advantage is that finite-rate chemistry is known to be a bottleneck for the whole hydrodynamic system. Improvements to the speed of execution of finite-rate chemistry within the context of a reactive CFD simulation will therefore have a significant and positive impact. Additionally, the thermochemical behavior of a reacting system is inherently constrained to a low-dimensional manifold, and machine learning is adept at capitalizing on low-dimensional intrinsic dimensionalities. There are also inviolable physical parameters that can be built into loss functions to ensure physical inference predictions from learned models.

In the literature, SciML methods are used to solve stiff ODEs effectively and efficiently. In the earlier work of Brown et al., \cite{brown2021novel}, ResNet was used to solve the stiff-ODEs governed by the chemically reacting flows. They used the reduced hydrogen-air reaction model with eight species and 18 reactions \cite{petersen1999reduced}. In \cite{ji2021stiff}, Ji et al. employed PINNs for stiff chemical kinetics. In \cite{kim2021stiff}, Kim et al. proposed novel derivative calculation techniques for the stability of stiff ODE systems, and they were employed with deep neural networks with proper scaling. In \cite{galaris2021numerical}, the authors proposed physics-informed random projection neural networks for solving the initial value problem for stiff ODE problems. For this purpose, the authors employed an extreme learning machine with one hidden layer network with radial basis functions. Similarly, De et al. \cite{de2022physics} employed PINNs with a theory of functional connections and extreme learning machines for stiff chemical kinetics. In \cite{anantharaman2021stably} Anantharaman et al. employed continuous-time Echo state networks for stiff quantitative systems pharmacology models. Recently, Zhang and Sankaran \cite{zhang2022autoencoder} proposed an autoencoder neural network-based reduced model to accelerate the simulation of chemical kinetic models with a large number of thermochemical state variables.

Our work develops procedures on how to learn solutions to stiff initial value problems using Deep Operator Networks (DeepONets), and how these algorithms are applied to finite rate chemical kinetics systems. DeepONets learn the stiff temporal evolution of chemical species’ mass fractions over a given duration during offline training, so that inference from the learned algorithm can evolve the thermochemical state during a prospective simulation at a rate comparable to the hydrodynamic time scale, but without sacrificing the fidelity of the chemical system’s transition path. Chemically reacting systems can frequently include many species that are not present for substantial periods of time, but that might only come into existence as products of one reaction and then immediately catalyze another reaction and be consumed just as quickly. These species are important to model but present a challenge to deep learning methods. We present results of using the DeepONet framework in several configurations to improve inference prediction accuracy with non-participatory species for stiff ODE systems, including finite rate kinetic models.

\section{Problem Statement}
\label{sec:problem_statement}
In this work, we consider four cases, the first case is the 3 species ROBER problem, the second case is the POLLU problem with 20 species, and the third and the fourth cases deal with a skeletal syngas mechanism for pure kinetics and turbulent flame, respectively. A brief discussion of the first two problems is provided alongside the results in Section ~\ref{sec:examples}.
In this section, we will provide context for problems $3$ and $4$. This work aims to learn the solution propagator for $\Delta t = \Delta t_{CFD}$, which is the time increment of advancing a CFD simulation for one iteration without the chemical source term, i.e. for passive transport. This time increment is dictated by the smallest flow time scale, which can be orders of magnitude larger than the chemistry time scales $\Delta t_{chm}$, i.e., $\Delta t_{CFD}\gg \Delta t_{chm}$. The chemistry time-advancement $\Delta t_{chm}$ is small either because of the difference between flow time scales and chemistry time scales, and/or because of  the stiffness of the chemical reactions. Consider the chemical kinetics equation:
\begin{equation}\label{eq:kin}
    \frac{d\Phi}{dt} = S(\Phi), 
\end{equation}
where $\Phi \in \mathbb{R}^{n_s+1}$ is the time-dependent vector of the thermochemical (TC) state of the system that contains the mass fraction of $n_s$ species and temperature and $S(\Phi): \mathbb{R}^{n_s+1} \longrightarrow  \mathbb{R}^{n_s+1}$ is the chemical source term. 
The solution propagator is denoted by $F_t^{t+\Delta t}: \mathbb{R}^{n_s+1} \longrightarrow  \mathbb{R}^{n_s+1}$, whose action can be defined as: 
 \begin{equation}
    \Phi(t+\Delta t) = F_t^{t+\Delta t} (\Phi(t)).
\end{equation}
Eq.~\ref{eq:kin} governs the evolution of $\Phi(t)$. The kinetics evolution equation is an autonomous dynamical system and therefore, it is possible to learn $ F_t^{t+\Delta t}$ irrespective of $t$. Adopting this viewpoint, $\Phi(t)$ for any $t$ represents an ``initial condition" and $F_t^{t+\Delta t}$ advances this initial condition to the next time step. We aim to learn $ F_t^{t+\Delta t}$ with $\Delta t =\Delta t_{CFD}$. Learning $ F_t^{t+\Delta t}$ can eliminate the stiffness issues imposed by Eq.~\ref{eq:kin} which is consequential for reactive flow simulations.

\section{Operator based frameworks}
\label{sec:approaches}

To establish an accurate description of reacting flows, detailed chemical kinetic models with a large number of chemical species must be incorporated, resulting in a complex and stiff system of ODEs and PDEs. This section will examine three methodologies for building a surrogate solution operator that is accurate and effective for chemical processes and interactions with fluid dynamics that involve several scales and significant nonlinearity. In the first methodology, all TC state variables are learned to map to the same states in advance using a neural operator called DeepONet \cite{lu2021learning}. In the second, we proposed a Partition-of-Unity (PoU)-based DeepONet method, where the evolving basis of the trunk net satisfies the property of PoU. Finally, in the third methodology, a compact latent representation of the chemical kinetics model is created using an autoencoder, and this representation is used to train the neural operator to improve the kinetics model over time. Details of the three methodologies are presented in this section.

\begin{figure}[htpb]
\centering
\includegraphics[trim=0cm 0cm 0cm 0cm,scale=0.8]{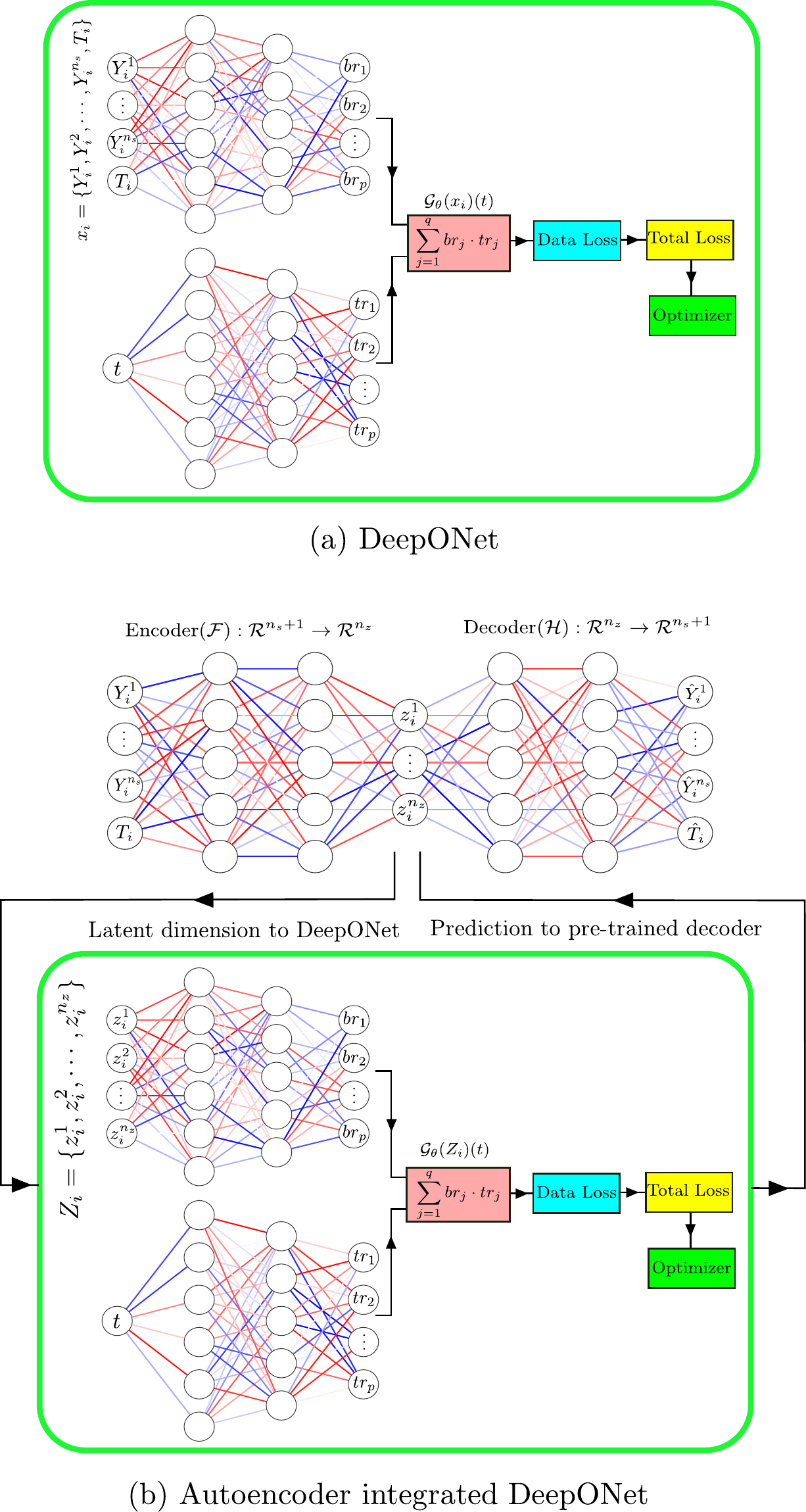}
\caption{(a) DeepONet: The DeepONet architecture consists of two deep neural networks (DNNs), namely the branch and the trunk networks. The branch network encodes the input function and comprises of concentration of species and temperature at a given time step. The trunk network encodes the information related to the temporal coordinates at which the solution operator is evaluated to compute the loss function. (b) Multi-layer autoencoder integrated DeepONet to obtain a compact latent representation of the chemical kinetics model for a given time step.}
\label{fig:frameworks}
\end{figure}

\subsection{Deep neural operator}
\label{subsec:deeponet}

Building quick simulators for solving parametric PDEs \cite{goswami2022physics} has become possible, thanks to the development of modern ML models. A new wave of techniques for speeding up the simulation of ODEs and PDEs is being proposed as the ML revolution continues to sweep the scientific community. Instead of discovering the PDE in an explicit form, it is usually important to have a substitute model of the PDE solution operator that can repeatedly simulate PDE solutions for different initial and boundary conditions. The neural operator, introduced in 2019 in the form of deep operator networks (DeepONet) \cite{lu2021learning}, which is inspired by the universal approximation theorem of operators, fulfills this promise. The theory that ensures a small approximation error (that is, the error between the target operator and the class of neural networks in a given finite-size architecture) has a particular influence on the architecture of DeepONet. 

A nonlinear continuous operator is used in operator regression to map one infinite-dimensional function to another. The DeepONet architecture consists of two deep neural networks (DNNs): the branch, and the trunk networks. The branch network encodes the input function, $\mathbf X$, comprising mass fractions, $Y$ of $n_s$ species and temperature, $T$ at a given time step, $t$. Specifically, $\mathbf X = \{\mathbf x_1, \mathbf x_2, \dots, \mathbf x_{N_s}\}$, where $N_s$ is the total number of samples, $\mathbf x_i = \{Y_i^1, Y_i^2, \dots, Y_i^{n_s}, T_i\}$, and $i\in \{1,N_s\}$. The trunk network encodes the information related to the temporal coordinates, $\zeta = \{t_{i}+\hat{t}_1, t_{i}+\hat{t}_2, \dots, t_{i}+\hat{t}_{n_t}\}$, at which the solution operator is evaluated to compute the loss function. Here, $n_t$ is the number of time steps. Consider an analytical or a computational model which simulates a physical process and represents a mapping between a vector of random variables, $\mathbf{X} \in \mathbb{R}^{D_{\text{in}}}$, and corresponding output quantities of interest (QoIs), $\mathbf{y}(\zeta) \in \mathbb{R}^{D_{\text{out}}}$ where $D_{\text{in}} = n_s+1$ and $D_{\text{out}} = n_t \times (n_s+1)$. The goal of the DeepONet is to learn the solution operator, $\mathcal G(\mathbf{X})$ that approximates the computational model and can be evaluated at continuous temporal coordinates, $\zeta$ (input to the trunk net). The output of the DeepONet for a specified input vector, $\mathbf{x}_i \in \mathbf{X}$, is a scalar-valued function of $\zeta$ expressed as $\mathcal G_{\boldsymbol\theta}(\mathbf{x}_i)(\zeta)$, where $\boldsymbol{\theta} = \left(\mathbf W, \mathbf b \right)$ includes the trainable parameters (weights, $\mathbf W$, and biases, $\mathbf b$) of the networks. In DeepONet \cite{lu2021learning}, the solution operator for an input realization, $\mathbf x_i$, is expressed as: 
\begin{equation}\label{eq:output1_deeponets}
    \begin{split}
      \mathcal G_{\boldsymbol \theta}(\mathbf{x}_i)(\zeta) &= \sum_{j = 1}^p br_j \cdot tr_j = \sum_{j = 1}^{p}br_j(\mathbf{x}_i)\cdot tr_j(\zeta),   
\end{split}
\end{equation}
where ${br_{1}, br_{2}, \ldots, br_p}$ are outputs of the branch network and ${tr_1, tr_2, \ldots, tr_p}$ are outputs of the trunk network. The trainable DeepONet parameters, denoted by $\boldsymbol{\theta}$ in Eq.~\eqref{eq:output1_deeponets} are typically optimized by minimizing a typical regression loss. Although the original DeepONet architecture proposed in \cite{lu2021learning} has shown remarkable success, several extensions have been proposed in \cite{lu2021comprehensive,kontolati2022influence} to modify its implementation and produce  efficient and robust architectures. A schematic representation of the original DeepONet architecture is shown in Fig.~\ref{fig:frameworks}(a).

\subsection{Partition-of-Unity (PoU) based deep neural operator}
In PoU-DeepONet, the Partition-of-Unity (PoU) constraint is enforced on the outputs of the trunk net where the evolving basis satisfies the $\sum_{j=1}^p tr_j = 1$ condition, where $tr_j$ are the outputs of the trunk net. The PoU is crucial for data interpolation. Additionally, PoU frequently allows extending local constructions to the whole space. The PoU-DeepONet is quite easy to implement. The loss function directly incorporates the PoU-based constraint as an additional penalty term.

\subsection{Autoencoder integrated deep neural operator}
\label{subsec:ae_deeponet}

Solving the governing equations that evolve a large number of TC state variables becomes challenging for real-world applications. These equations also often involve multiple state variables, such as temperature, pressure, density, and species concentrations, which interact in complex ways. Hence, the solution to these equations often involves trade-offs between accuracy, computational efficiency, and simplicity. To accelerate such simulations, we present an autoencoder (AE)-based DeepONet approach in this section. An AE is a prominent nonlinear dimensionality reduction method that allows for increased compressibility of nonlinear data. As a result, it is extremely appealing for highly nonlinear chemical reaction systems. With AE, a more compact latent representation can be achieved to mitigate the negative effects induced by the acceptable and prevalent circumstance that many species' mass fractions are zero during combustion system integration. The fact that most of the potential TC states in a combustion system reside on or near a lower-dimensional manifold in spite of the vast number of species in detailed chemical kinetic models serves as the impetus for using an AE to resolve the highly nonlinear chemical kinetics. 

In this work, we have employed multi-layer autoencoders (AE) \cite{oommen2022learning} to obtain a compact latent representation of the chemical kinetics model for a given time step, $t$. Essentially, the AE defines the mapping between the full system, $\mathbf{x}_i$ and a reduced system $\mathbf{z}_i = \{z_i^1,z_i^2,...,z_i^{n_z}\}$, \textit{i.e.}, $\mathcal{F}: \mathcal{R}^{n_s+1} \rightarrow \mathcal{R}^{n_z}$ and $\mathcal{H}: \mathcal{R}^{n_z} \rightarrow \mathcal{R}^{n_s+1}$, where $n_z \ll n_s+1$. In AE, the encoding and decoding mappings are obtained via self-supervised learning. An encoder and a decoder, which are represented by the mappings $\mathcal{F}$ and $\mathcal{G}$, respectively, make up the two components of the AE. The encoder takes the TC state vector $\mathbf{X}$ as input and compresses it into a much smaller vector $\mathbf{Z}$ at the latent layer, where $\mathbf{z}_i \in \mathbf{Z}$. The compressed vector $\mathbf{Z}$ is referred to as the latent representation of $\mathbf{X}$. The decoder takes the latent representation $\mathbf{Z}$ as input and reconstructs or maps it back to the original space, $\mathbf{\hat{X}}$, where $\mathbf{\hat{x}} \in \mathbf{\hat{X}}$. The AE learns the representation by finding the optimal weights that minimize the difference between the output, i.e., reconstructed dataset $\mathbf{\hat{X}}$ and the input, \textit{i.e.}, the original dataset $\mathbf{X}$ composed of a large collection of TC state samples. To find a low-dimensional latent representation of the TC states, the AE is initially trained using a common regression loss. This makes it possible to trace the development of a chemical system by solving the initial TC equations and the equations of the latent representation. The DeepONet is then trained to learn the evolving kinetics of the latent space in the following step. To this end, the DeepONet learns the mapping, $\mathcal G_{\boldsymbol \theta}: \mathcal{R}^{n_z} \rightarrow \mathcal{R}^{n_t\times n_z}$, where the input to the branch network is the latent representation at time $t$, while the trunk network inputs $\zeta = \{t_{i}+250, t_{i}+500, t_{i}+750, t_{i}+1000\}$, and $n_t= 4$. We deploy the DeepONet for unobserved test samples after it has been trained and its parameters optimized. To obtain a representation of all the TC states at the time steps taken into account by $\zeta$, the output of the DeepONet is used as input to the decoder, $\mathcal{H}$. A schematic representation of the AE-integrated DeepONet architecture is shown in Fig.~\ref{fig:frameworks}(b).

\section{Results}
\label{sec:examples}

In this section, we discuss four examples that make use of the three DeepONet architectures we covered in Section~\ref{sec:approaches}. The \texttt{TensorFlow} library \cite{agarwal2016ten} has been used for the implementation. Tables \ref{table:architectures} and \ref{table:architectures-MLAE} contain a list of the architectures employed for learning the dynamics of the problems discussed in this section.

\begin{table}[h!]
\caption{DeepONet architectures for all four examples.}
\centering
\begin{tabular}{c c c c}
\toprule
 Model & Branch net & Trunk net & \begin{tabular}{@{}c@{}}Activation \\ function\end{tabular}  \\
 \toprule
Example $1$  & $[1, 120, 120, 120, 100]$ & $[1, 120, 120, 120, 120, 100]$ & Sine (Adaptive \cite{jagtap2020adaptive,jagtap2020locally}) \\
Example $2$  & $[1, 120, 120, 120, 100]$ & $[1, 120, 120, 120, 120, 100]$ & Sine (Adaptive \cite{jagtap2020adaptive,jagtap2020locally}) \\  
Example $3$(a)  &  $[14, 128, 128, 128, 120]$ & $[1, 128, 128, 128, 120]$ & Tanh \\
Example $3$(b)  &  $[l_d\small{+}2, 128, 128, 128, 10\small{\times}l_d]$ & $[1, 128, 128, 128, 10\small{\times}l_d]$ & Leaky ReLU \\
Example $4$(a)  &  $[12, 128, 128, 128, 120]$ & $[1, 128, 128, 128, 120]$ & Tanh \\
Example $4$(b)  &  $[l_d, 128, 128, 10\small{\times}l_d]$ & $[1, 128, 128, 10\small{\times}l_d]$ & Tanh \\
\bottomrule
\multicolumn{4}{l}{(a) denotes the architecture for learning dynamics using the DeepONet framework.}\\
\multicolumn{4}{l}{(b) denotes the architecture for learning dynamics using the Autoencoder-integrated DeepONet framework.}
\end{tabular}
\label{table:architectures}
\end{table}

\begin{table}[ht!]
\begin{center}
\caption{Architecture of multi-layer autoencoders (MLAE).}
\begin{tabular}{ c c c c } 
\hline
Application & MLAE & $l_d$ & \begin{tabular}{@{}c@{}}Activation \\ function\end{tabular}\\
\hline
Example $3$ & $[12, 128, 128, 128, l_d, 128, 128, 128, 12]$ & $6$ & Leaky ReLU  \\ 
Example $4$ & $[10, 256, 64, l_d, 64, 128, 10]$ & $2$ & Leaky ReLU  \\ 
\hline
\end{tabular}
\label{table:architectures-MLAE}
\end{center}
\end{table}

\subsection{Example 1: ROBER problem}
\label{subsec:example1}

\begin{figure} [htpb] 
\centering 
\includegraphics[trim=0cm 0cm 0cm 0cm,scale=0.34]{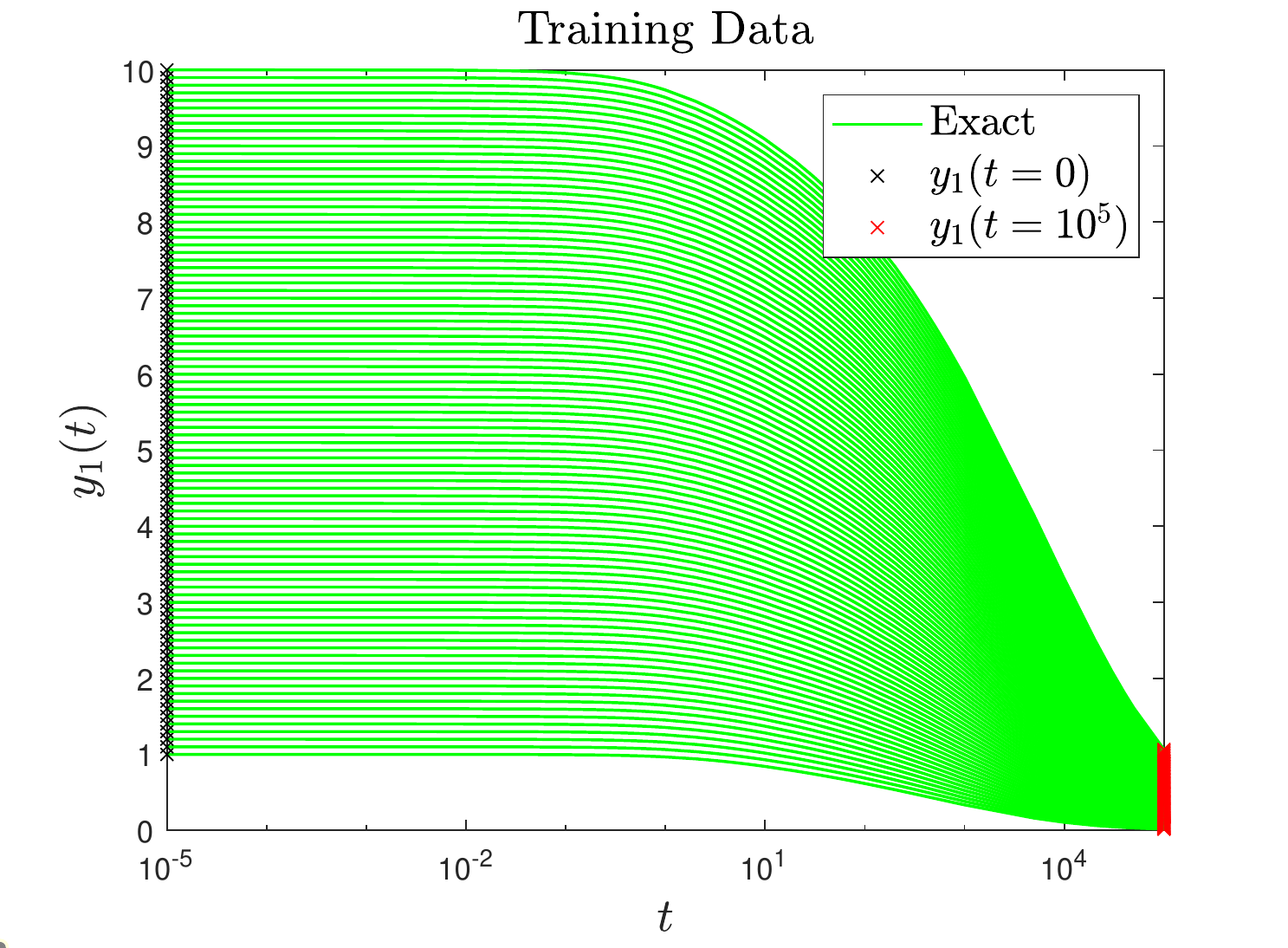}
\includegraphics[trim=0cm 0cm 0cm 0cm,scale=0.34]{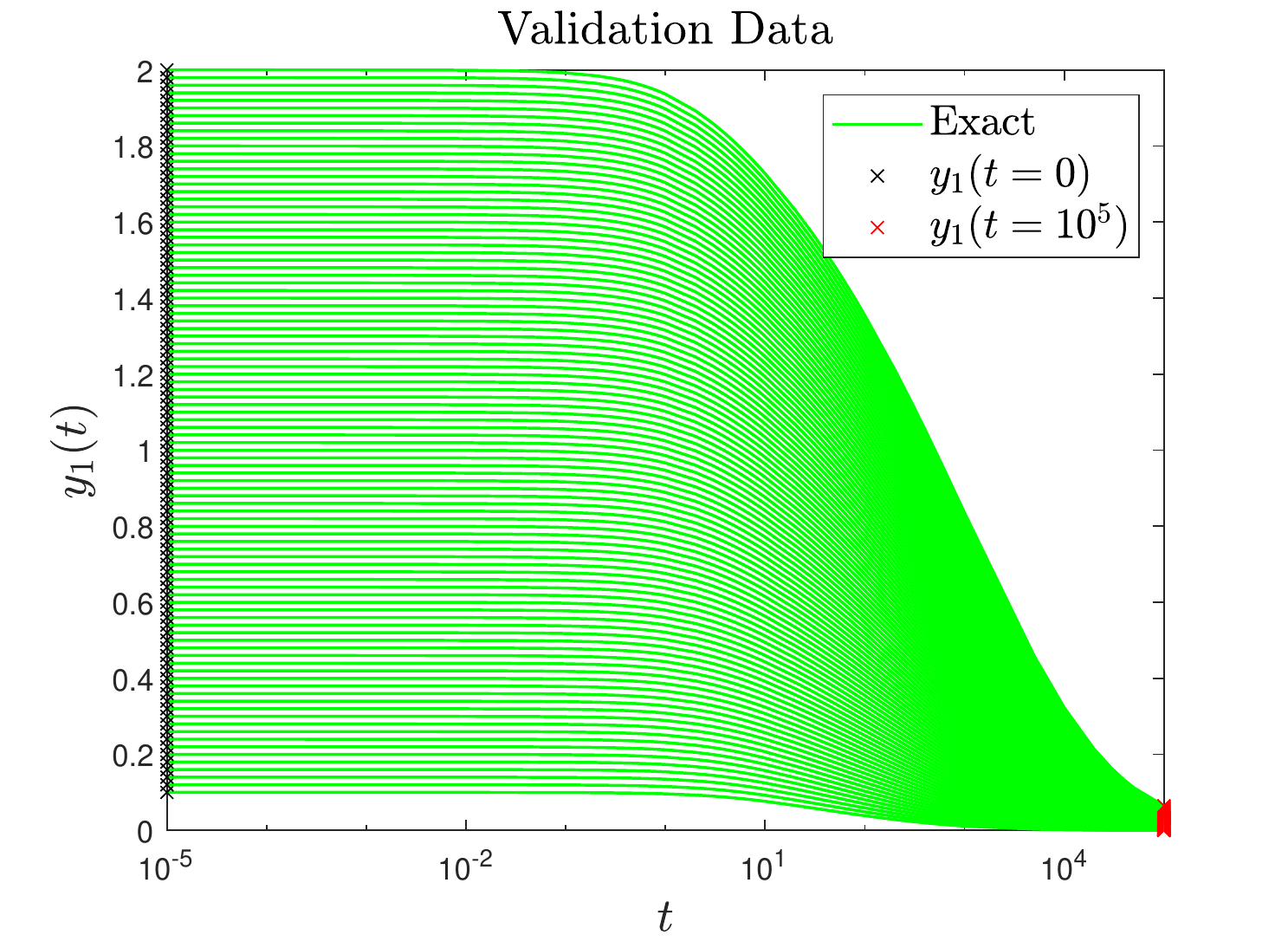}
\includegraphics[trim=0cm 0cm 0cm 0cm,scale=0.34]{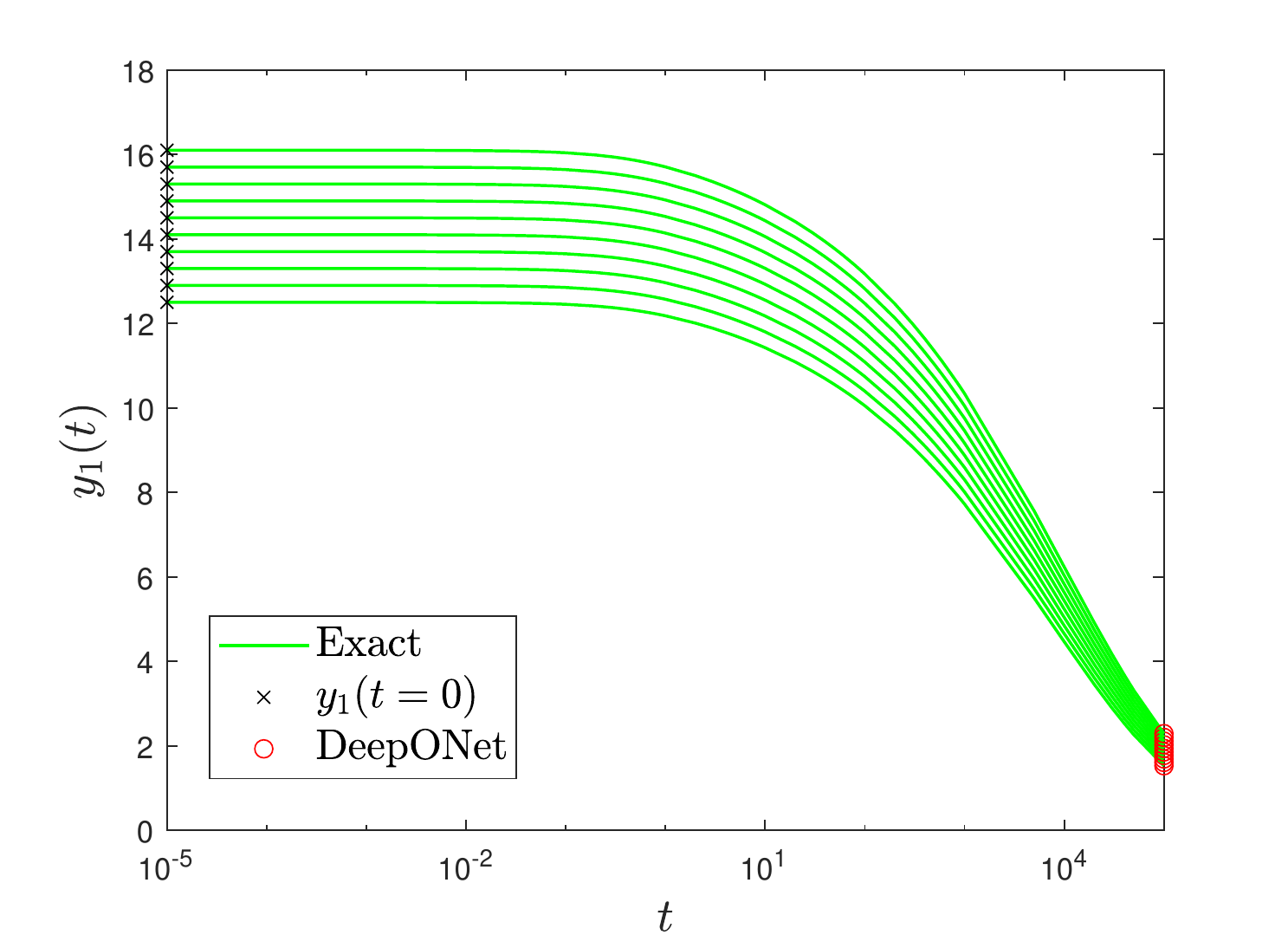}
\includegraphics[trim=0cm 0cm 0cm 0cm,scale=0.34]{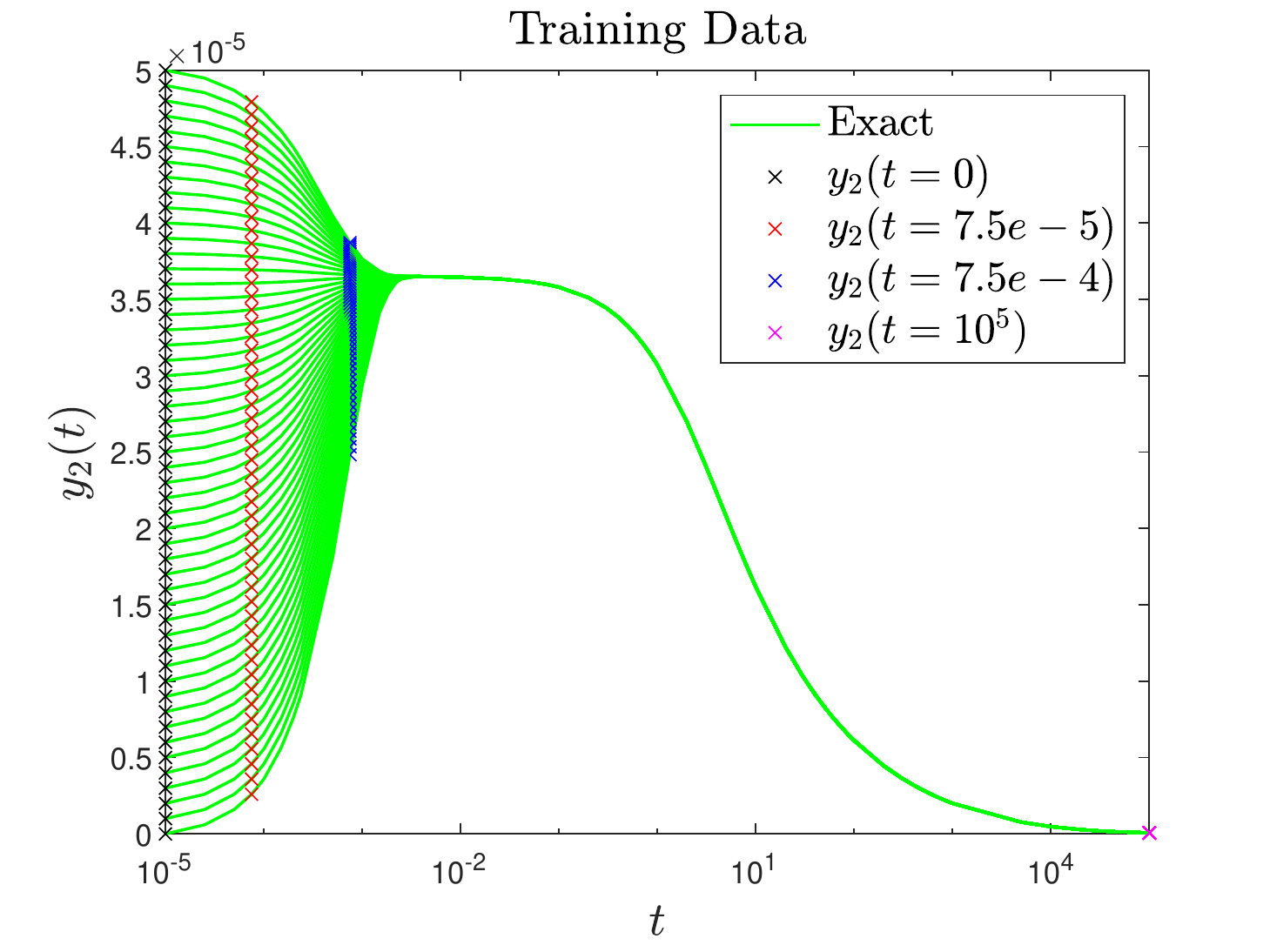}
\includegraphics[trim=0cm 0cm 0cm 0cm,scale=0.34]{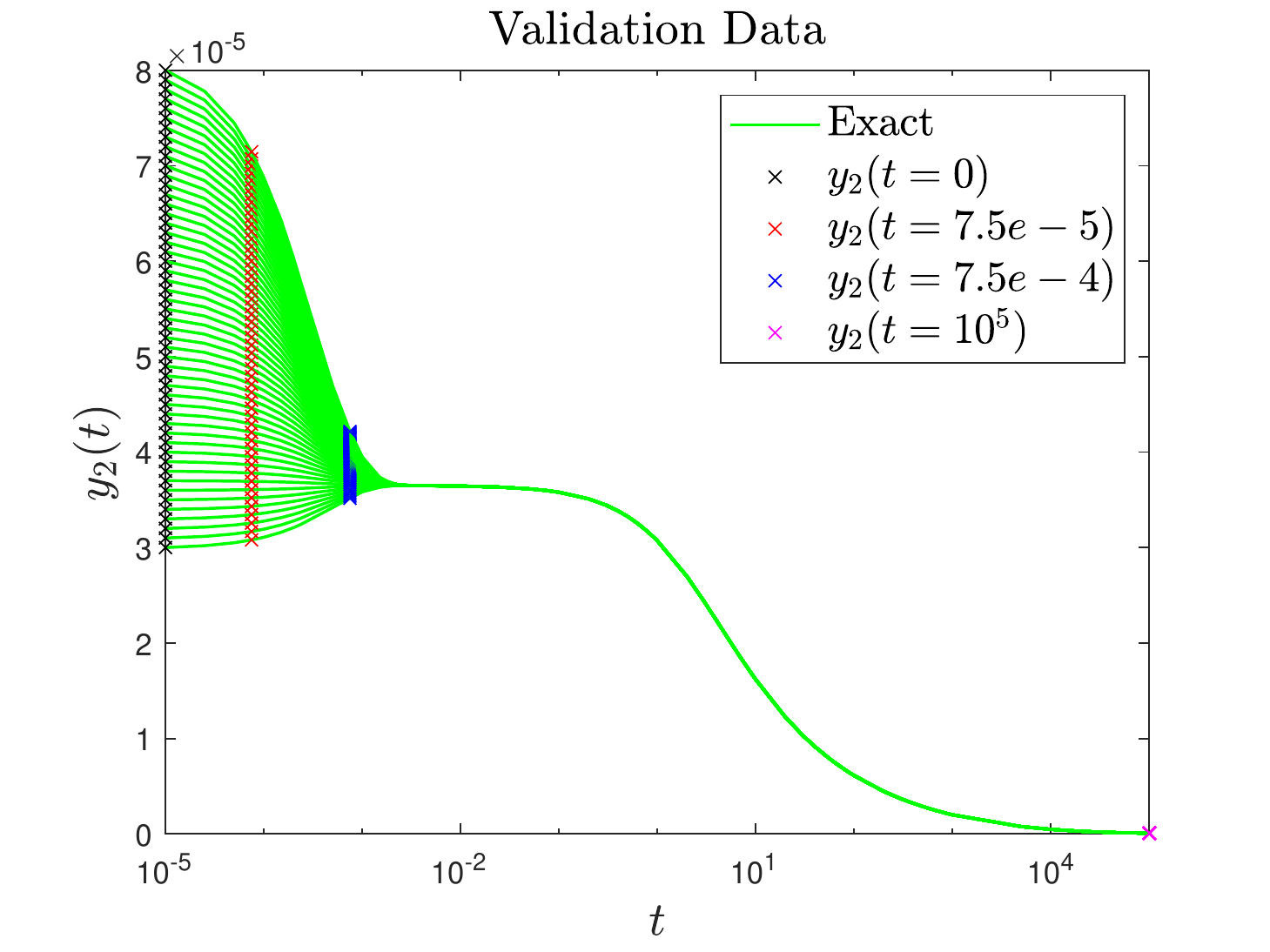}
\includegraphics[trim=0cm 0cm 0cm 0cm,scale=0.34]{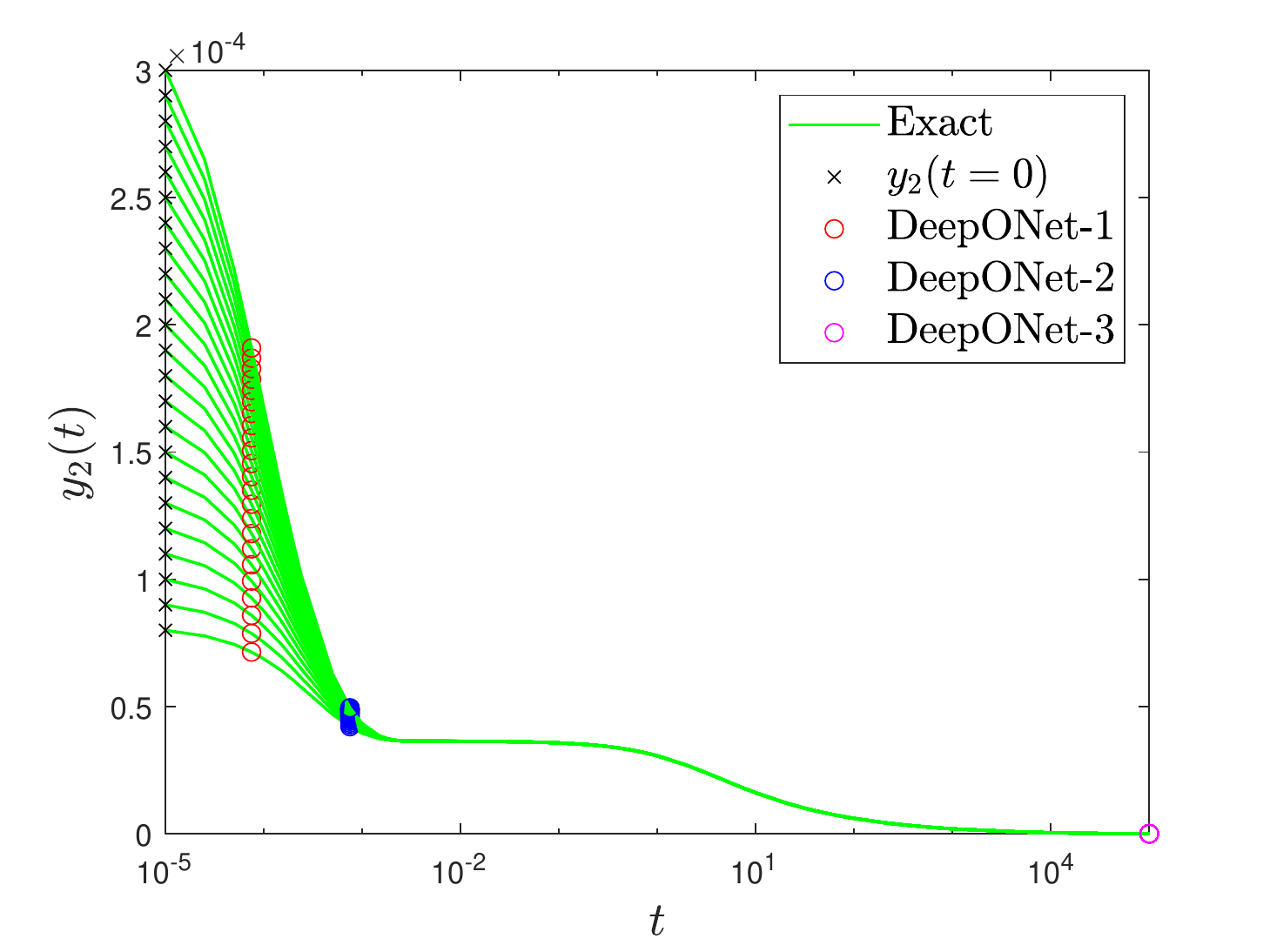}
\includegraphics[trim=0cm 0cm 0cm 0cm,scale=0.34]{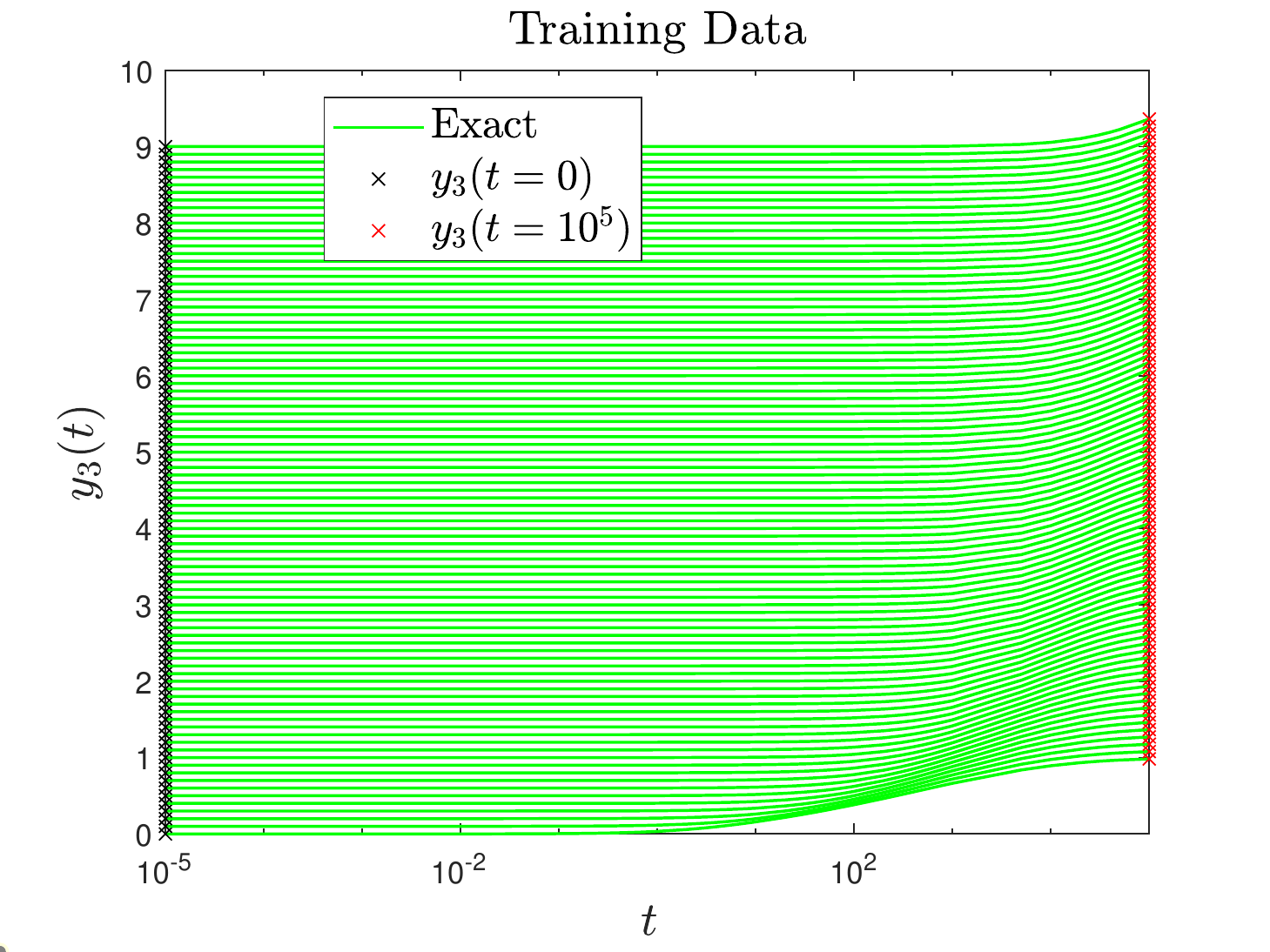}
\includegraphics[trim=0cm 0cm 0cm 0cm,scale=0.34]{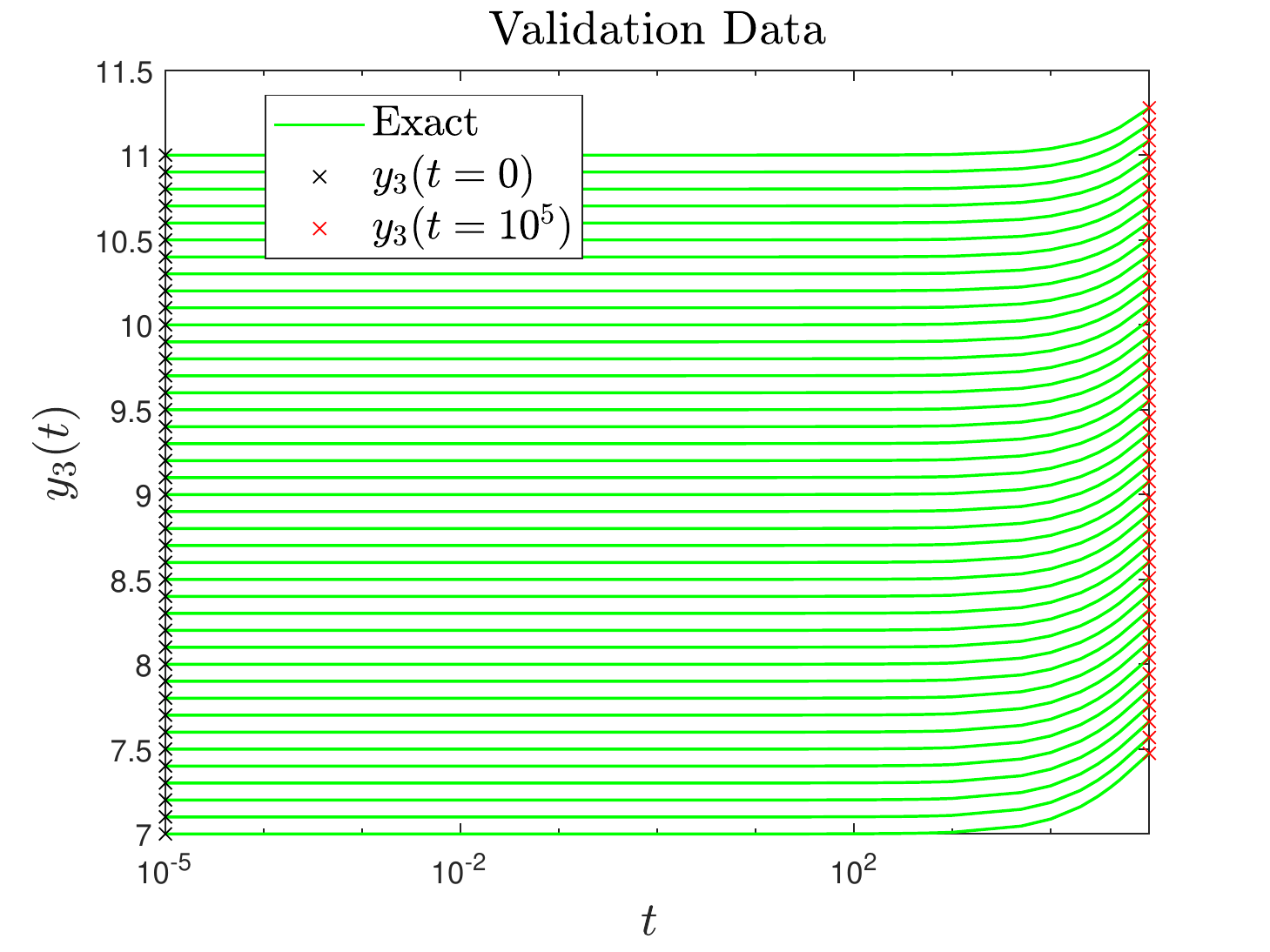}
\includegraphics[trim=0cm 0cm 0cm 0cm,scale=0.34]{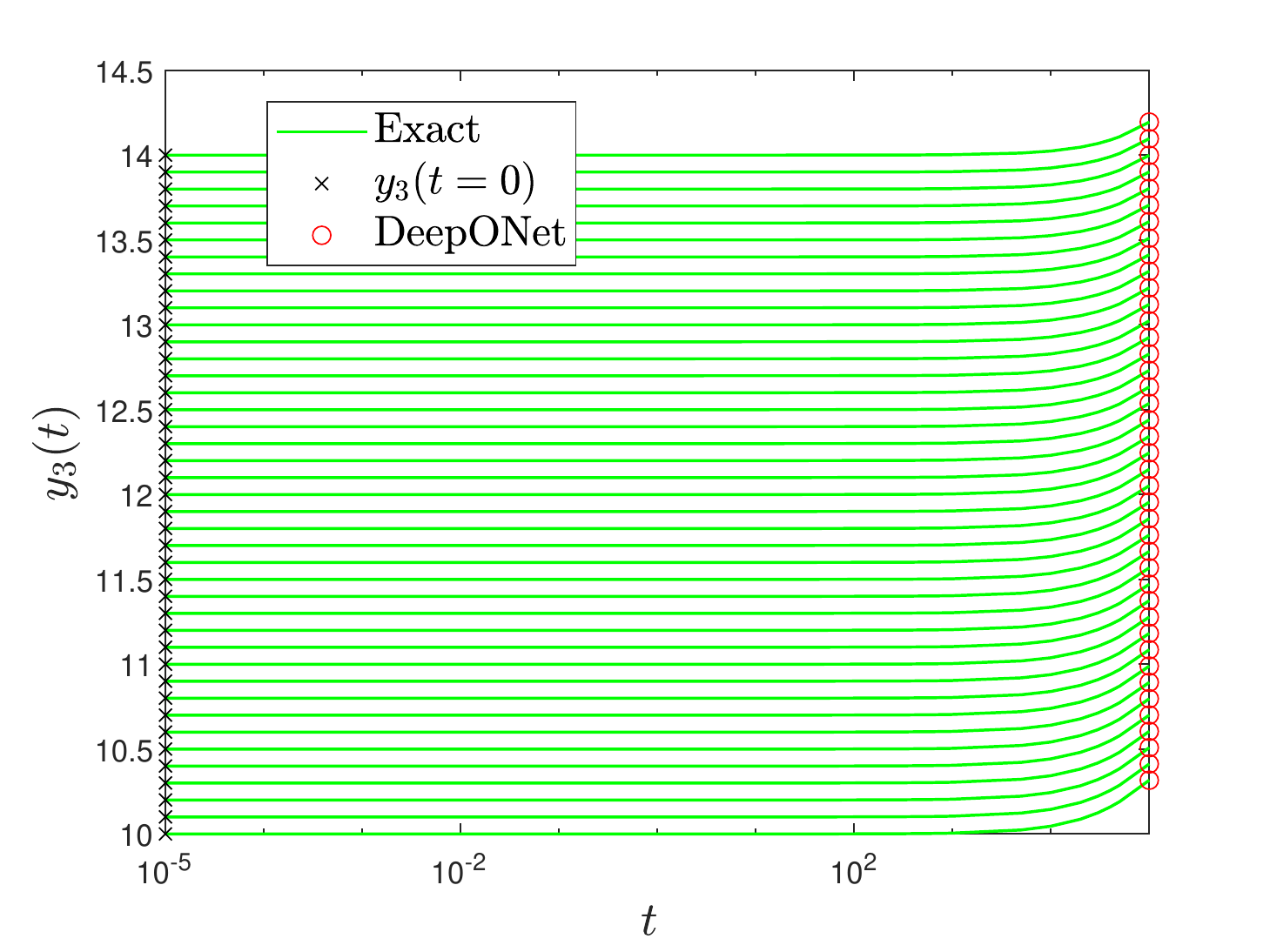}
\caption{ROBER problem: Training data is shown in the left column, where the input and output datasets for species $y_1, y_2$ and $y_3$ are shown in black and red $\times$, respectively. The green line shows the exact solution for the ROBER system. Similarly, the testing data is shown in the middle, and the corresponding PoU-DeepONet result for extrapolated data is shown in the right column for species $y_1, y_2$ and $y_3$.}
\label{fig:ROB1}
\end{figure}

Robertson’s equations, denoted as ROBER, are one of the prominent stiff systems of ODEs. It is given by the following reaction network:
\begin{align}
A  & \xrightarrow{k_1} B, \\
B+ B  & \xrightarrow{k_2} C + B, \\
B+ C  & \xrightarrow{k_3} A + C.
\end{align}
The reaction rate constants are given by $k_1 = 0.04, k_2 = 3 \times 10^7 , k_3 = 10^4$, and the initial conditions are given by $y_1 (0) = 1, y_2 (0) = 0, y_3 (0) = 0$, where $y_1, y_2, y_3$ denote the concentrations of the species $A, B, C$, respectively. The evolution of the species concentrations can be described by the following ODEs:
\begin{align}\label{robers}
\frac{dy_1}{dt} &= -k_1y_1 + k_3y_2 y_3, \\
\frac{dy_2}{dt} &= k_1y_1 - k_2y_2^2 - k_3y_2 y_3, \\
\frac{dy_3}{dt} &= k_2y_2^2.
\end{align}
In this section, we shall solve the ROBER problem using PoU-DeepONet. The PoU constraint is enforced on the outputs of the trunk net where the formed basis satisfies the $\sum_{j=1}^p tr_j = 1$ condition. The input/output dataset is generated for three species $y_1 , y_2, y_3$ by parameterizing them. The data is divided into training and testing (validation). We used a sine activation function that is locally adaptive \cite{jagtap2020adaptive,jagtap2020locally}. Rowdy activations \cite{jagtap2022deep} can also be used to accelerate the training, especially when a multi-modal dataset is involved. The comprehensive survey about the activation functions is given in \cite{jagtap2022important}. Once the PoU-DeepONet is trained over these data sets, we use extrapolated input data to infer the output. Figure \ref{fig:ROB1} shows the training and validation data sets for $y_1, y_2$ and $y_3$ in the first two columns, and the corresponding PoU-DeepONet result (last column) for extrapolated data, respectively. 

Table \ref{Table1} shows the number of training, validation, and extrapolation data sets along with their domain range for all the species, $y_1, y_2$, and $y_3$. The maximum point-wise errors with mean and standard deviation (for 10 different realizations) are given in the last row. 
\begin{table}[htpb]
\centering
\caption{ROBER problem: The details of training, validation (testing), and extrapolation data sets for the three species, $y_1, y_2$, and $y_3$. The corresponding maximum point-wise errors with mean and standard deviation (for $10$ different realizations) are shown in the last row. Note that, each species is approximated by individual PoU-DeepONet with the same architecture shown in table \ref{table:architectures}.}
\resizebox{\columnwidth}{!}{%
\begin{tabular}{|c|c|c|c|c|c|}
\hline
\multirow{3}{*}{}  &\multirow{3}{*}{$\mathbf{y_1}$} & \multicolumn{3}{c|}{$\mathbf{y_2}$} & \multirow{3}{*}{$\mathbf{y_3}$}  %
\\ \cline{3-5}
 & & DeepONet-1& DeepONet-2& DeepONet-3& \\
\hline
 $\#$Training Data & 901 & 601 & 601 & 601 & 901\\
 (Range) & ($1\leq y_1 \leq 10$) &($0\leq y_2 \leq 5e-5$)  & ($2.5e-6\leq y_2 \leq 4.85e-5$) & ($2.5e-5\leq y_2 \leq 4e-5$)&  ($0\leq y_3\leq 9$)\\
\hline
 $\#$ Validation Data& 401 & 401 & 401 & 401 & 301 \\
 (Range) &  ($0.1\leq y_1 \leq 2$)&($3e-5\leq y_2 \leq 8e-5$)  & ($3.2e-6\leq y_2 \leq 7.2e-5$) & ($3.5e-5\leq y_2 \leq 4.5e-5$)& ($7\leq y_3\leq 11$)\\
\hline
 $\#$ Extraploation Data& 10 & 23 & 23 & 23 & 40 \\
 (Range) & ($12\leq y_1 \leq 16$) &($0.8e-4\leq y_2 \leq 3e-4$)  & ($0.7e-4\leq y_2 \leq 1.9e-4$) & ($0.4e-4\leq y_2 \leq 0.5e-4$)& ($10\leq y_3\leq 14$)\\
\hline
 & & & & & \\
Max. Point-wise error& 2.386e-2 $\pm$ 5.563e-3 & 5.752e-3 $\pm$ 7.462e-4 & 2.573e-3 $\pm$ 2.753e-4 & 1.543e-2  $\pm$ 6.930e-4 & 4.75e-2  $\pm$ 3.753e-3\\
 (Mean \& Std. Deviation) & & & & & \\
\hline
\end{tabular}}

\label{Table1}
\end{table}

\subsection{Example 2: POLLU problem}
\label{subsec:example2}

\begin{figure} [htpb] 
\centering 
\includegraphics[trim=0.0cm 0cm 1cm 0cm,scale=0.37]{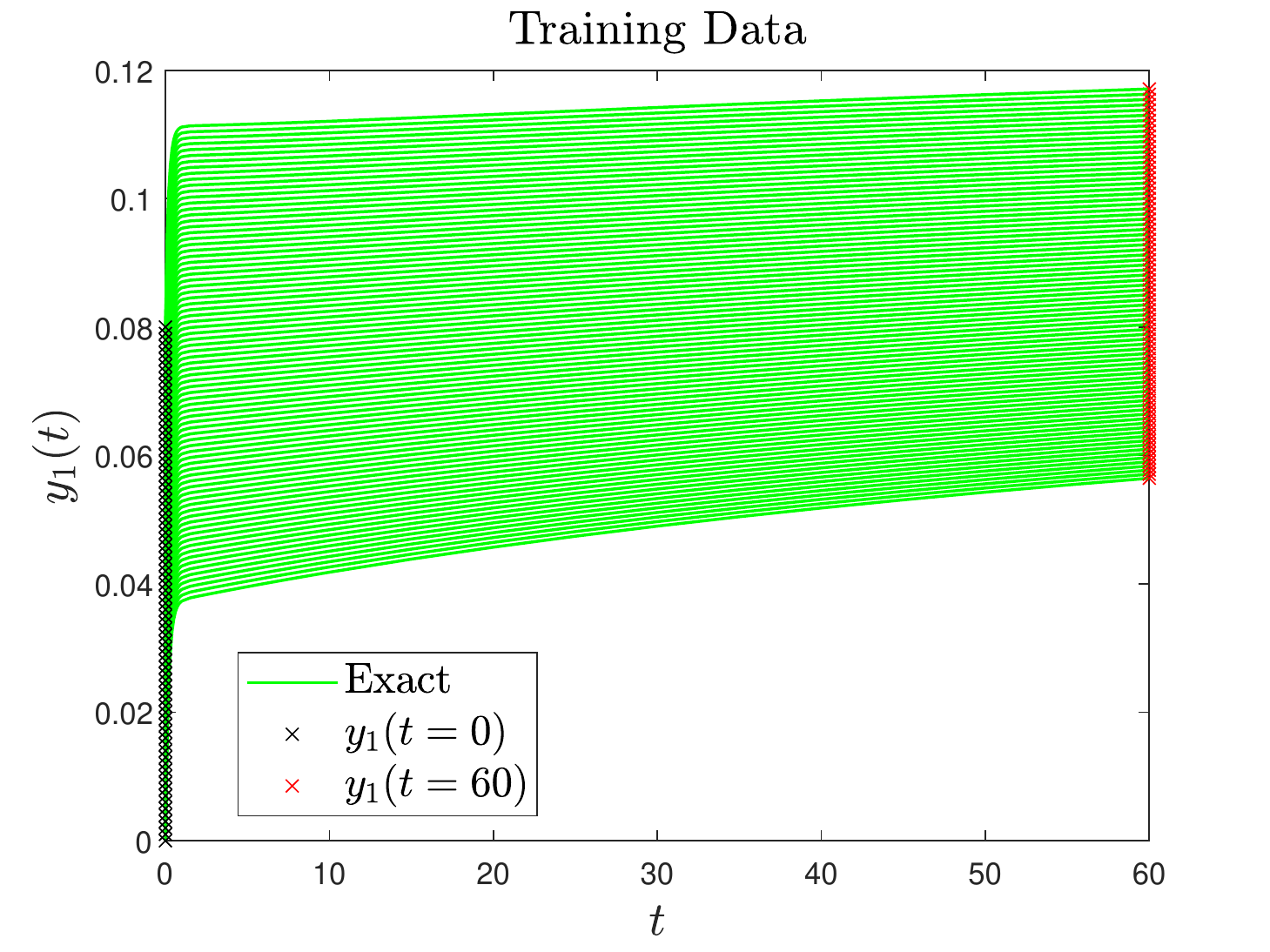}
\includegraphics[trim=0cm 0cm 1cm 0cm,scale=0.37]{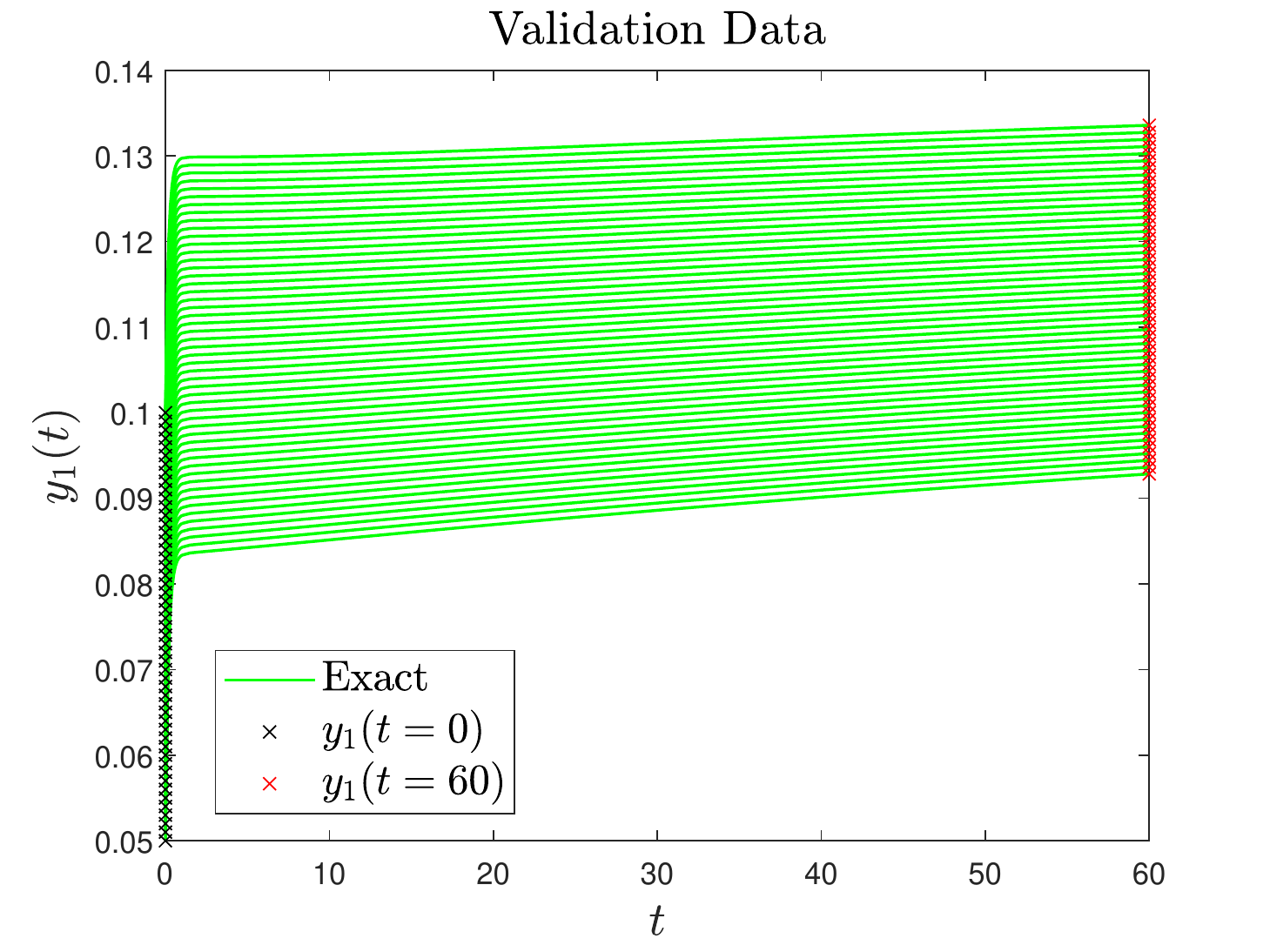}
\includegraphics[trim=0.0cm 0cm 1cm 0cm,scale=0.37]{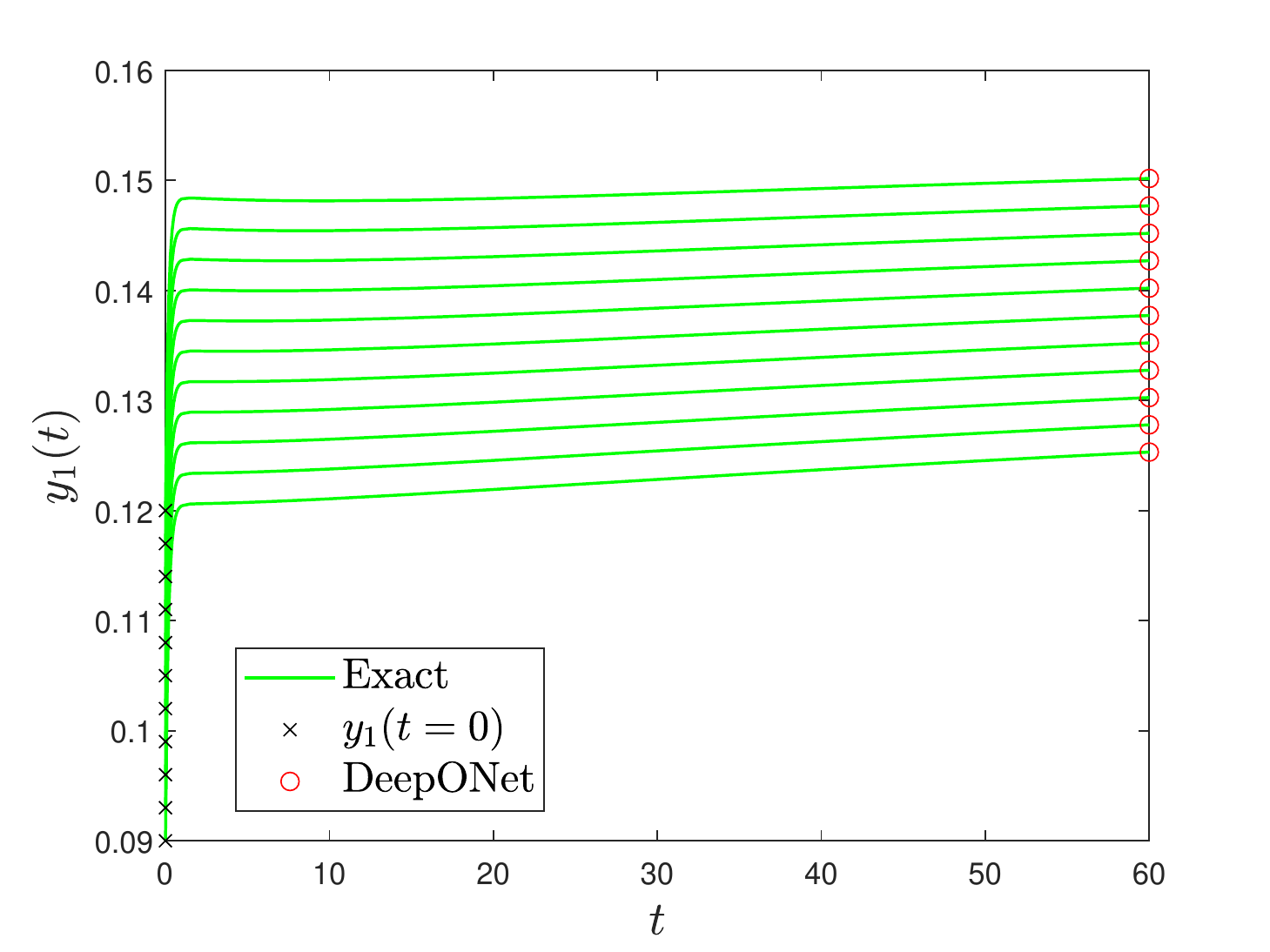}
\includegraphics[trim=0.0cm 0cm 1cm 0cm,scale=0.37]{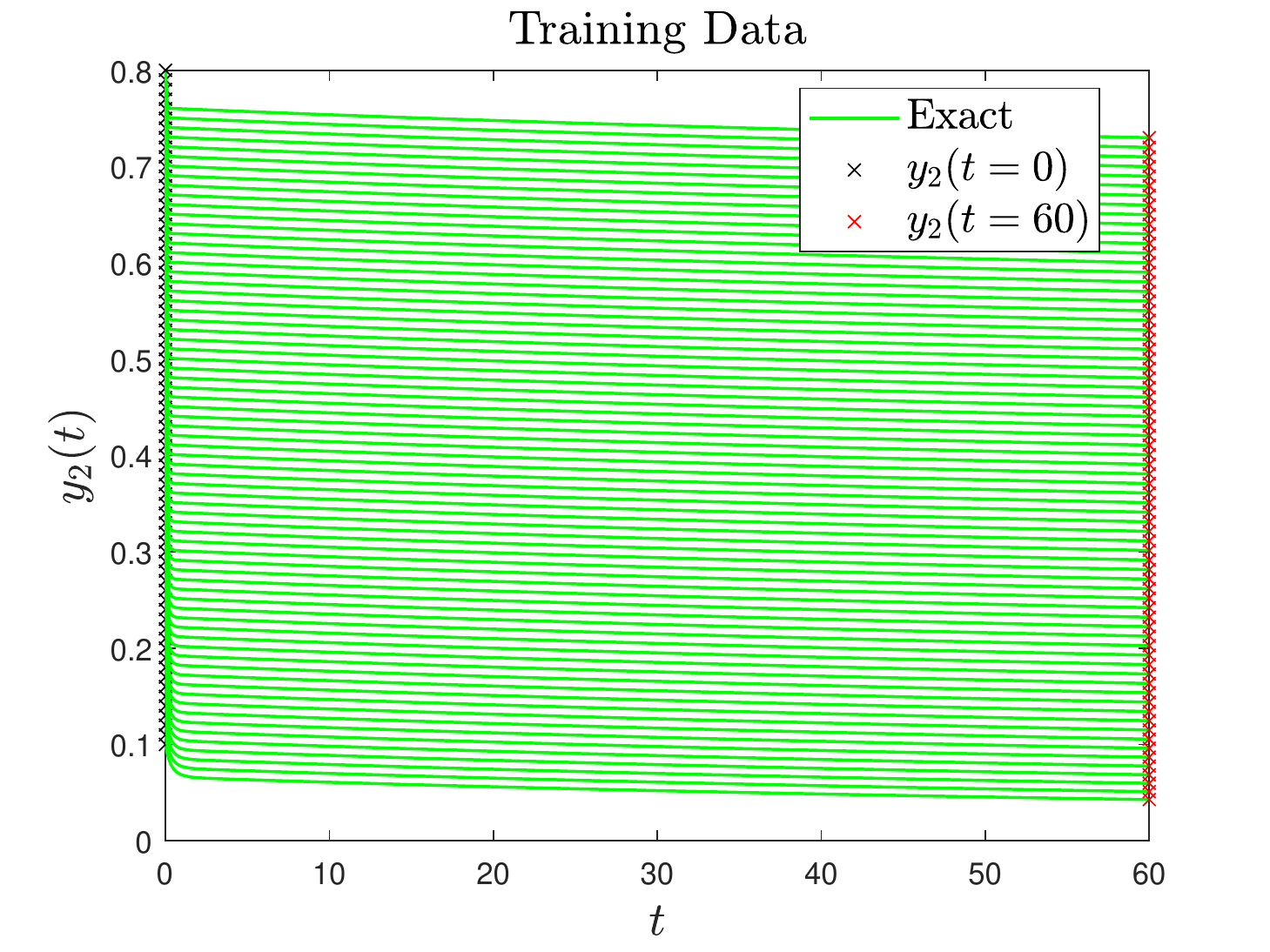}
\includegraphics[trim=0cm 0cm 1cm 0cm,scale=0.37]{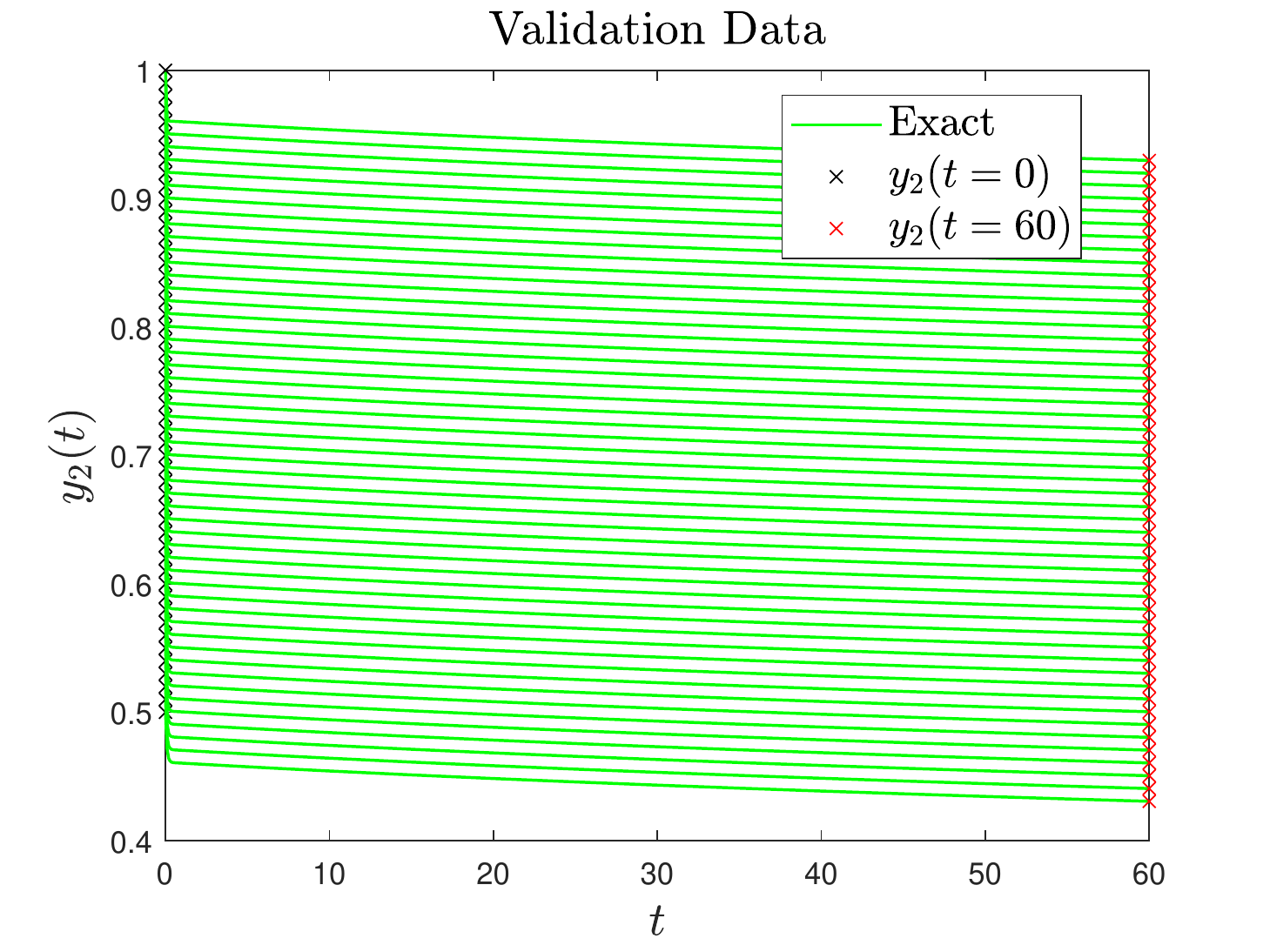}
\includegraphics[trim=0.0cm 0cm 1cm 0cm,scale=0.37]{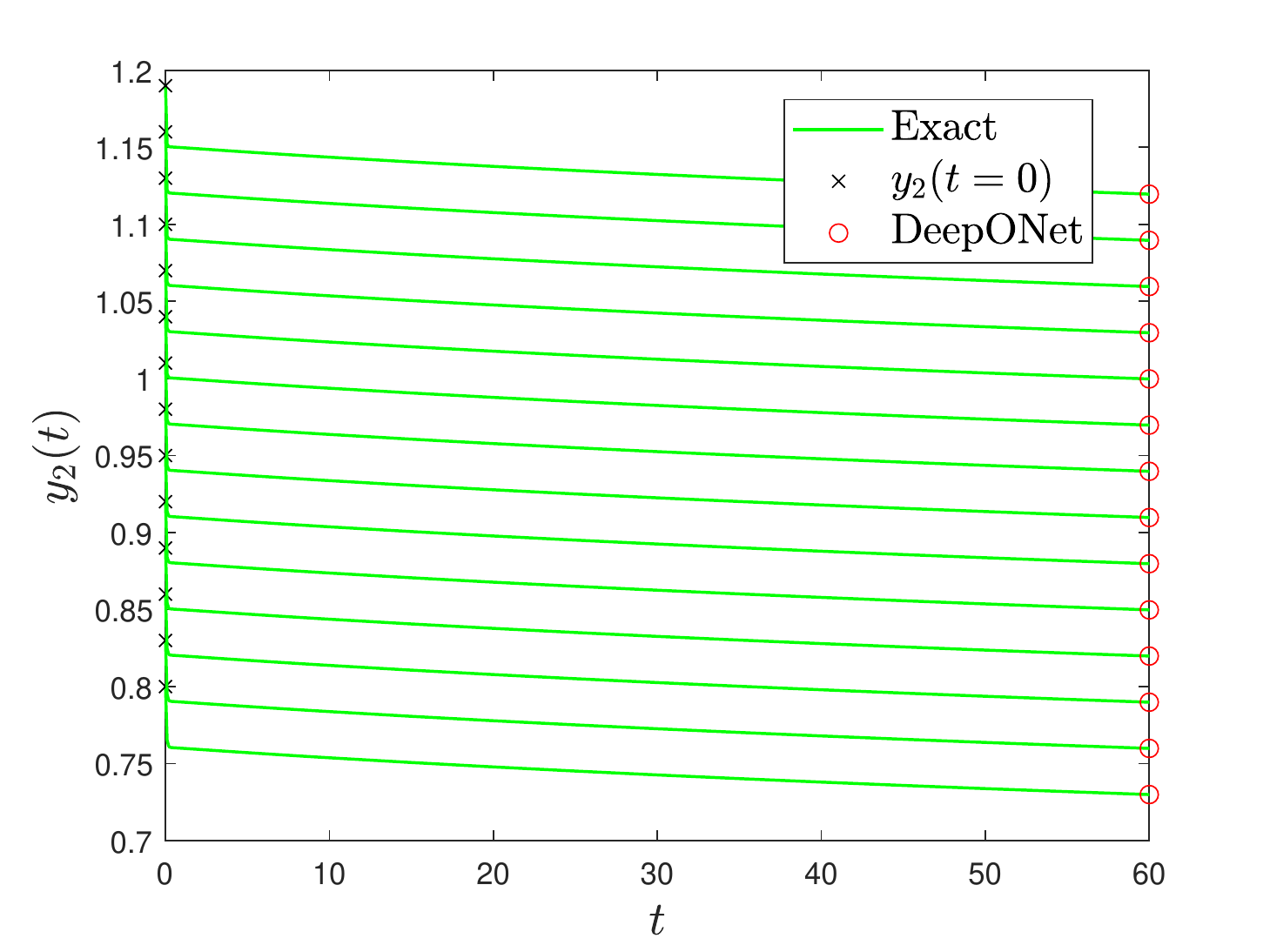}
\includegraphics[trim=0.0cm 0cm 1cm 0cm,scale=0.37]{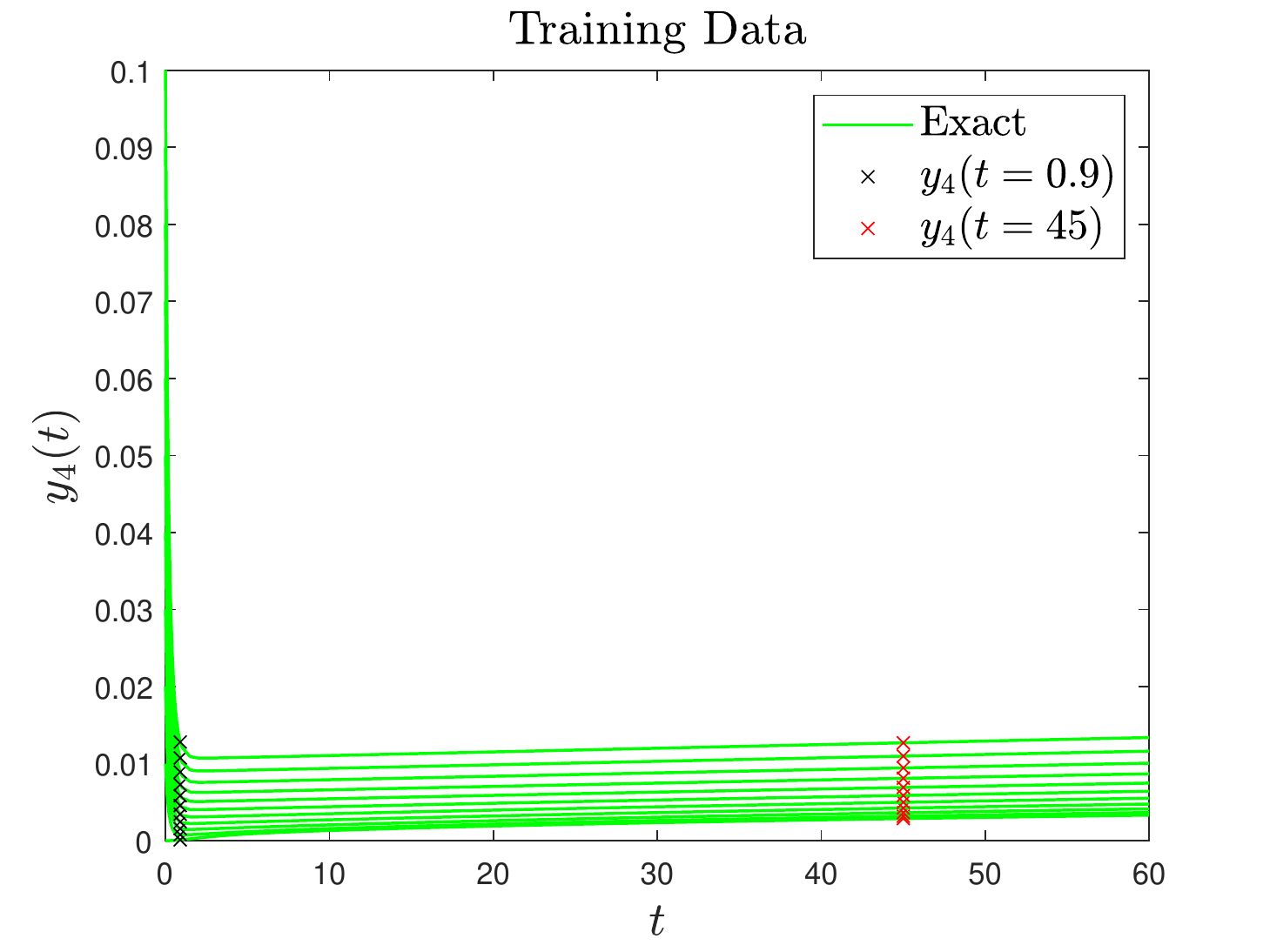}
\includegraphics[trim=0cm 0cm 1cm 0cm,scale=0.37]{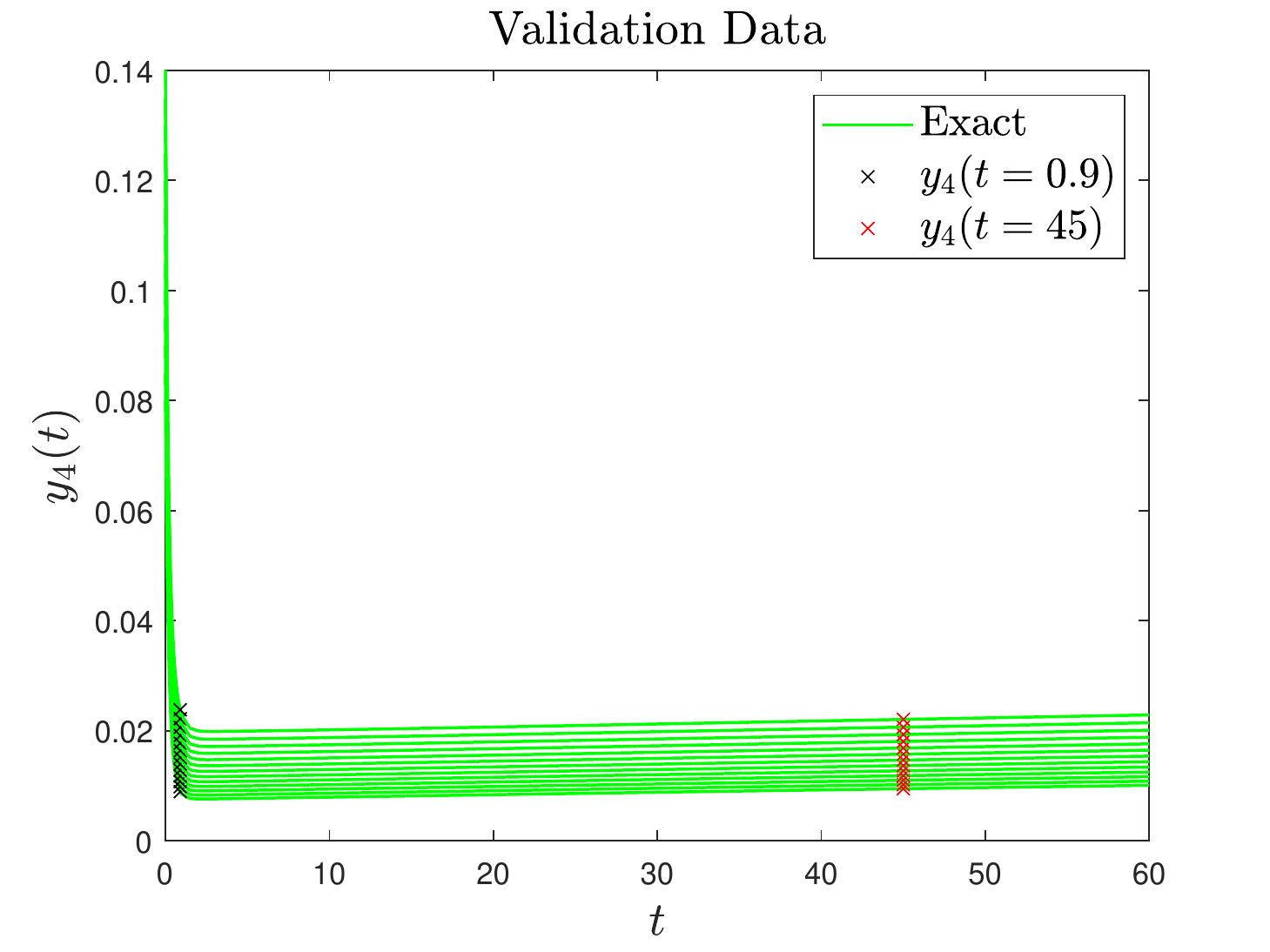}
\includegraphics[trim=0.0cm 0cm 1cm 0cm,scale=0.37]{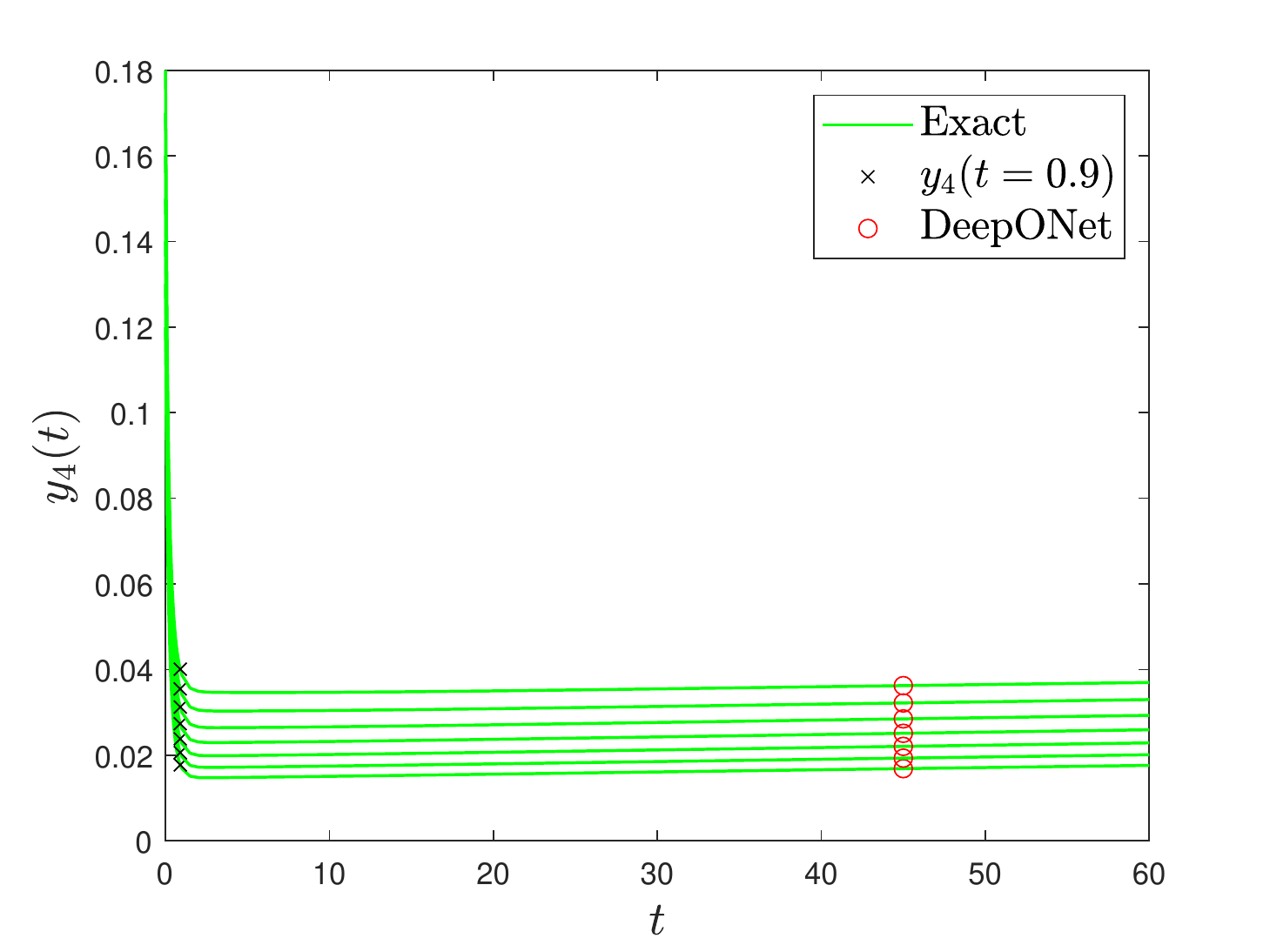}
\includegraphics[trim=0.0cm 0cm 1cm 0cm,scale=0.37]{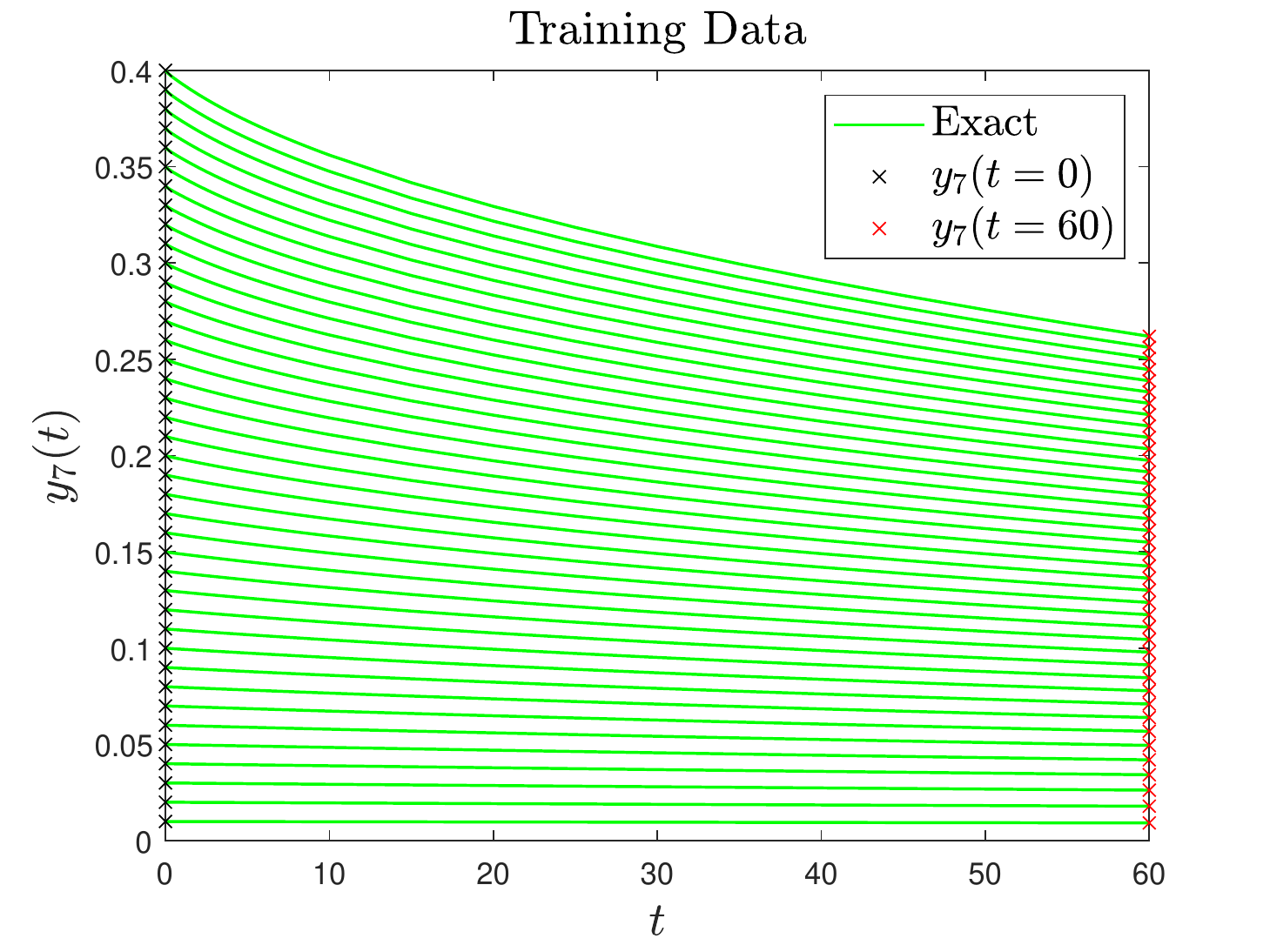}
\includegraphics[trim=0cm 0cm 1cm 0cm,scale=0.37]{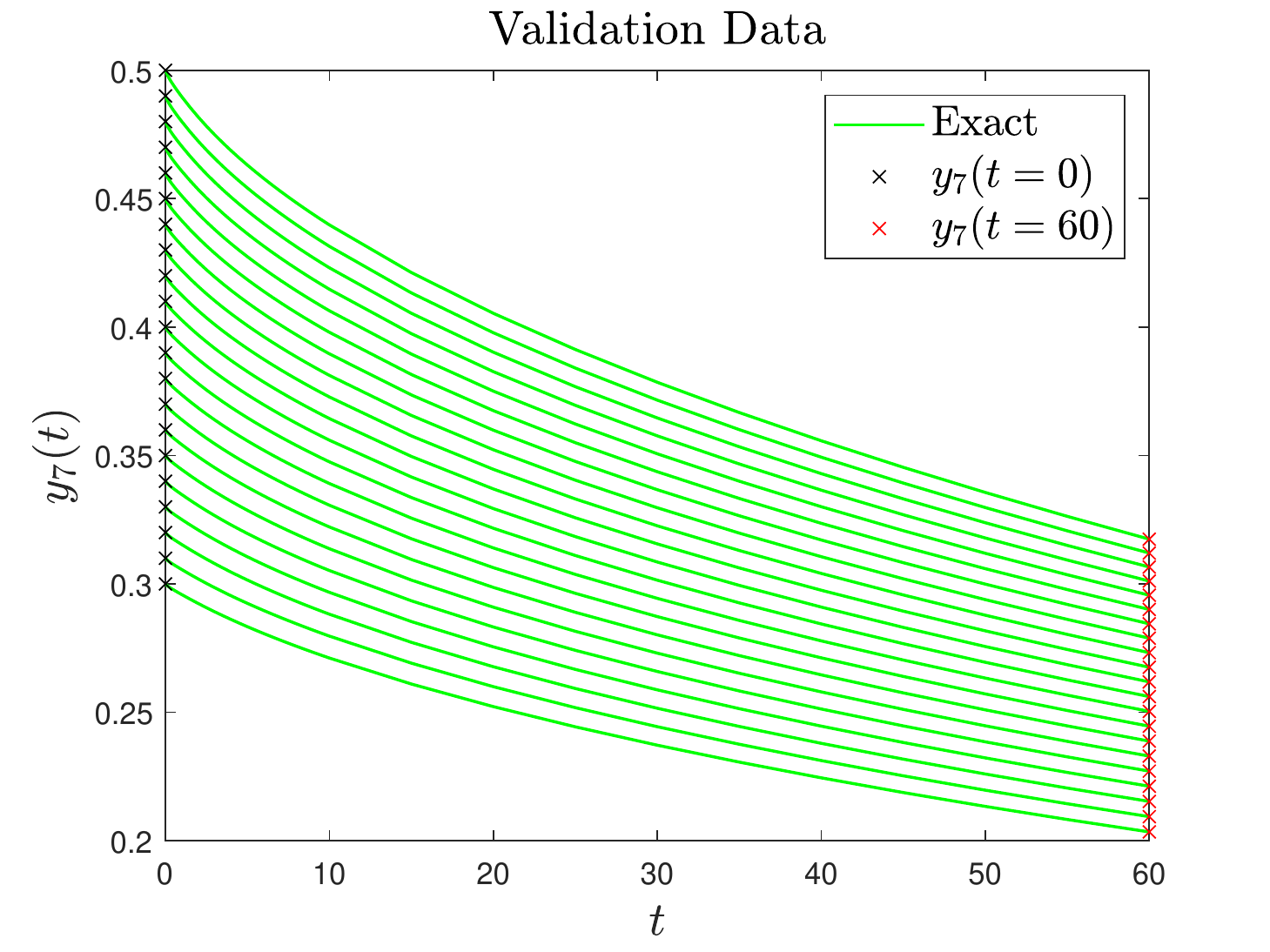}
\includegraphics[trim=0.0cm 0cm 1cm 0cm,scale=0.37]{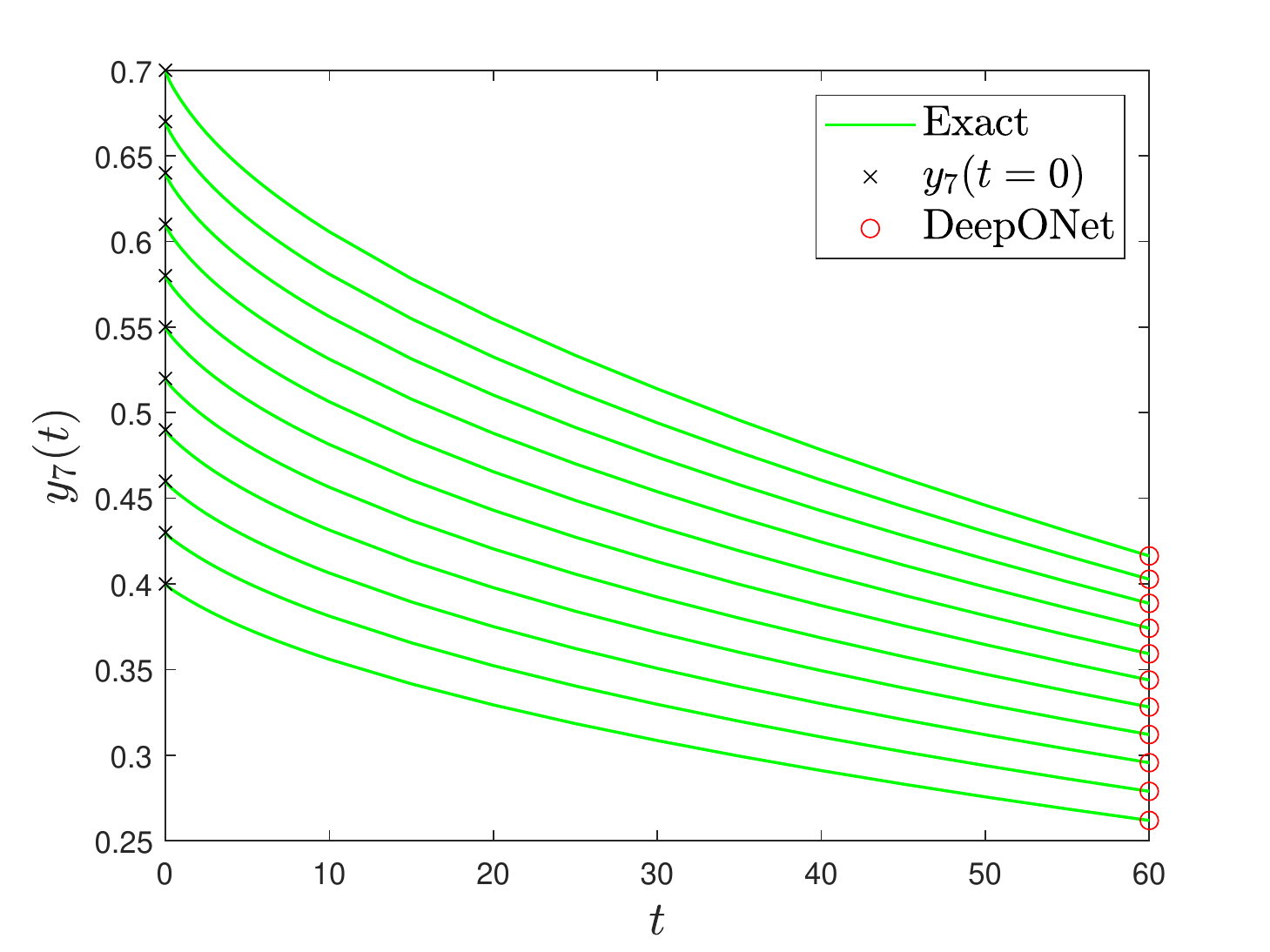}
\includegraphics[trim=0.0cm 0cm 1cm 0cm,scale=0.37]{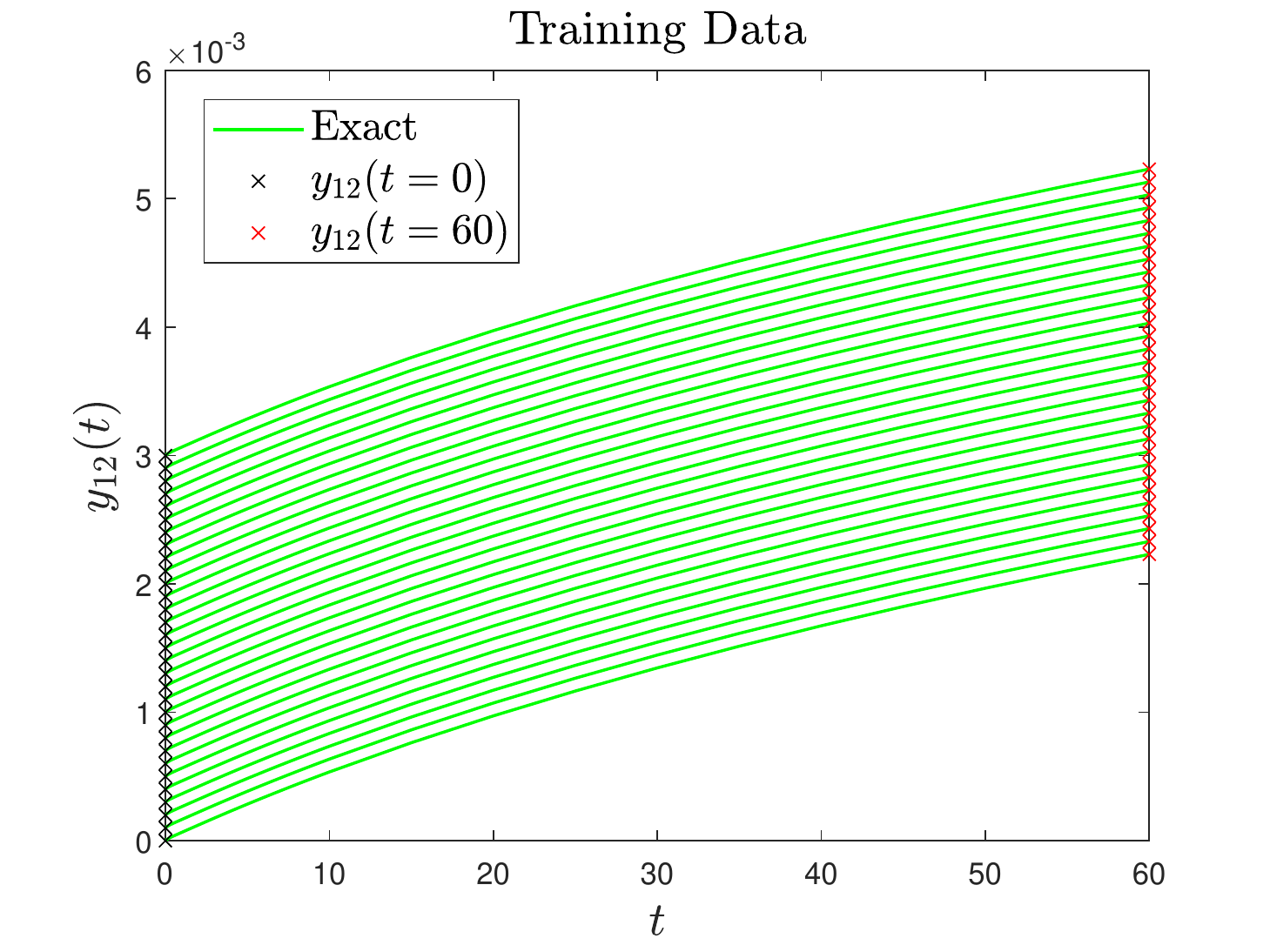}
\includegraphics[trim=0cm 0cm 1cm 0cm,scale=0.37]{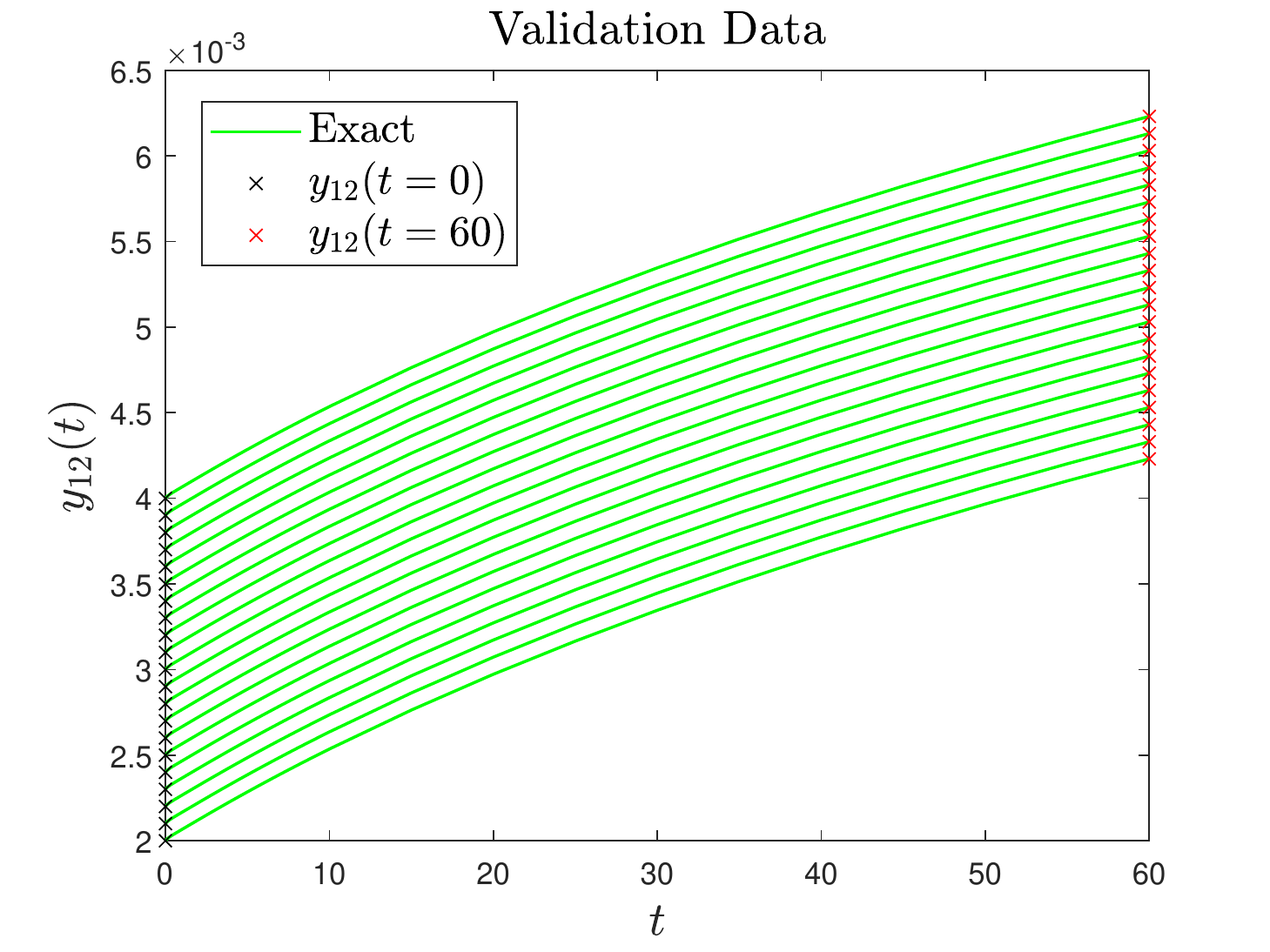}
\includegraphics[trim=0.0cm 0cm 1cm 0cm,scale=0.37]{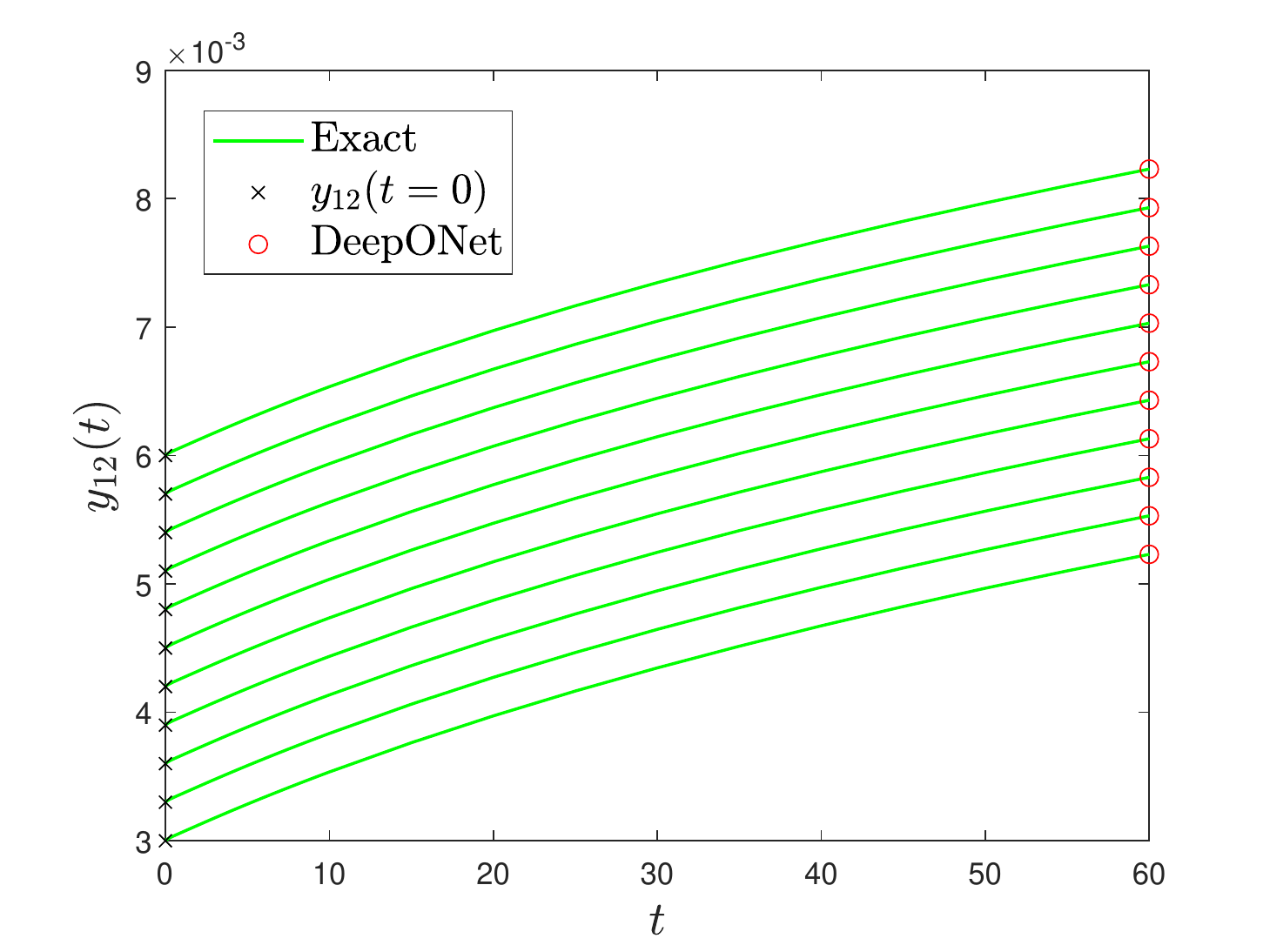}
\caption{POLLU problem: Training data is shown in the left column where the input and output dataset for species $y_1, y_2, y_4, y_7$ and $y_{12}$ are shown in black and red $\times$, respectively. The green line shows the exact solution for the POLLU system. Similarly, the testing data is shown in the middle and the corresponding PoU-DeepONet result for extrapolated data is shown in the right column. }
\label{fig:POL1}
\end{figure}

POLLU is an air pollution model developed at the Dutch National Institute of Public Health and Environmental Protection. It has $20$ species and $25$ reactions, given by
\begin{equation}
 \frac{dy}{dt} = g(y(t)), ~ y(0) = y_0,~  y\in\mathbb{R}^{20}, t>0,
\end{equation}
where
\begin{align*}
 y & = [y_1,y_2,\cdots,y_{20}],~~~ y_0 = [0, 0.2, 0, 0.04, 0, 0, 0.1, 0.3, 0.01, 0, 0, 0, 0, 0, 0, 0, 0.007, 0, 0, 0], \\
g(y(t)) & = \begin{bmatrix}
-r_1 - r_{10} - r_{14} - r_{23} - r_{24} + r_2 + r_3 + r_9 + r_{11} + r_{12} + r_{22} + r_{25}\\
-r_2 -r_3 - r_9 - r_{12} + r_1 + r_{21} \\
-r_{15} + r_1 + r_{17} + r_{19} + r_{22}\\
-r_2 - r_{16} - r_{17}- r_{23} + r_{15}\\
-r_3 + 2r_4 + r_6 + r_7 + r_{13} + r_{20}\\
-r_6 - r_8 - r_{14} - r_{20} + r_3 + 2r_{18}\\
-r_4 - r_5 - r_6 + r_{13}\\
r_4 + r_5 + r_6 + r_7\\
-r_7 - r_8\\
-r_{12} + r_7 + r_9\\
-r_9 - r_{10} + r_8 + r_{11}\\
r_9\\
-r_{11} + r_{10}\\
-r_{13} + r_{12}\\
r_{14}\\
-r_{18} - r_{19} + r_{16}\\
-r_{20}\\
r_{20}\\
-r_{21} - r_{22} - r_{24} + r_{23} + r_{25}\\
-r_{25} + r_{24}
\end{bmatrix},
\end{align*}
and the reaction rates for all the species is given in table \ref{table2}.
\begin{table}[htpb]
\caption{Reaction rates for POLLU model} \label{table2}
\begin{center}
\begin{tabular}{ |c|c|c|c|c|c| } \hline
Reaction & r     & k & Reaction & r     & k\\ \hline
1   & $k_1 y_1$           &.350E+00 & 14  & $k_{14} y_1 y_6$    &.163E+05\\
2   & $k_2 y_2 y_4  $     &.266E+02& 15  & $k_{15} y_3   $     &.480E+07\\
3   & $k_3 y_5 y_2 $      &.120E+05& 16  & $k_{16} y_4  $      &.350E-03\\
4   & $k_4 y_7 $          &.860E-03& 17  & $k_{17} y_4  $      &.175E-01\\
5   & $k 5 y 7$           &.820E-03& 18  & $k_{18} y_{16}$     &.100E+09\\
6   & $k_6 y_7 y_6$       &.150E+05& 19  & $k_{19} y_{16} $    &.444E+12\\
7   & $k_7 y_9 $          &.130E-03& 20  & $k_{20} y_{17} y_6$ &.124E+04\\
8   & $k_8 y_9 y_6$       &.240E+05& 21  & $k_{21} y_{19}  $   &.210E+01\\
9   & $k_9 y_{11} y_2 $   &.165E+05& 22  & $k_{22} y_{19} $    &.578E+01\\
10  & $k_{10} y_{11} y_1$ &.900E+04& 23  & $k_{23} y_1 y_4 $   &.474E-01\\
11  & $k_{11} y_{13} $    &.220E-01& 24  & $k_{24} y_{19} y_1$ &.178E+04\\
12  & $k_{12} y_{10} y_2 $&.120E+05& 25  & $k_{25} y_{20} $    &.312E+01 \\
13  & $k_{13} y_{14} $    &.188E+01 &&&\\
 \hline
\end{tabular}
\end{center}
\end{table}
The first two columns of Figure \ref{fig:POL1} show the training and validation datasets for $y_1, y_2, y_4$, $y_7$, and $y_{12}$, and the corresponding PoU-DeepONet result (last column) for extrapolated data.
Table \ref{table3} shows the number of training, validation, and extrapolation datasets along with their domain range for all the species $y_i, i = 1,\cdots, 12$. The mean and standard deviation of the maximum point-wise error ($10$ different realizations) in the PoU-DeepONet results are given in the last column. In all cases, PoU-DeepONet predictions are accurate for the extrapolated data.

\begin{table}[htpb]
\begin{center}
\caption{Traning, validation (testing), and extrapolation data sets for POLLU problem using PoU-DeepONet. The corresponding maximum point-wise errors in terms of mean and standard deviation (for 10 different realizations) are shown in the last column. Note that, each species is approximated by individual PoU-DeepONet with the same architecture shown in table \ref{table:architectures}.}
\begin{tabular}{ |c|c|c|c|c| } \hline
 Species & $\#$Training Data    & $\#$Validation Data & $\#$Extrapolation Data & Max. Point-wise error \\ 
  & (Range)    & (Range)  & (Range)  & (Mean \& Std. Deviation)  \\ \hline
$y_1$&801  & 351 & 11 & \\
& ($0 \leq y_1 \leq 0.08)$  & ($0.05 \leq y_1 \leq 0.1$)  & ($0.09 \leq y_1 \leq 0.12$)  & 2.692e-3 $\pm$ 6.493e-4  \\ \hline
$y_2$& 501 & 301 & 14 & \\
& ($0.1 \leq y_2 \leq 0.8$)  & ($0.5 \leq y_2 \leq 1$)  & ($0.8 \leq y_2 \leq 1.2$)  & 1.893e-2 $\pm$ 9.426e-4\\ \hline
$y_{3}$& 601 & 331 & 12 & \\
& ($0.0 \leq y_3 \leq 0.4$)  & ($0.25 \leq y_3 \leq 0.4$)  & ($0.35 \leq y_3 \leq 0.6$)  &2.129e-2 $\pm$ 2.480e-3 \\ \hline
$y_4$& 401 & 201 & 6 & \\
& ($0.0 \leq y_4 \leq 0.1$)  & ($0.3 \leq y_4 \leq 0.5$)  & ($0.12 \leq y_4 \leq 0.18$)  & 4.495e-3 $\pm$ 7.426e-4\\ \hline
$y_{5}$& 341 & 251 & 10 & \\
& ($0.0 \leq y_5 \leq 0.5$)  & ($0.35 \leq y_5 \leq 0.55$)  & ($0.45 \leq y_5 \leq 0.7$)  & 1.495e-2 $\pm$ 2.006e-3\\ \hline
$y_{6}$& 391 & 221 & 8 & \\
& ($0.01 \leq y_6 \leq 0.4$)  & ($0.3 \leq y_6 \leq 0.5$)  & ($0.4 \leq y_6 \leq 0.7$)  & 5.484e-3 $\pm$ 4.727e-4\\ 
 \hline
 $y_{7}$& 401 & 301 & 20 &  \\
& ($0.2 \leq y_7 \leq 0.3$)  & ($0.2 \leq y_7 \leq 0.35$)  & ($0.3 \leq y_7 \leq 0.45$)  & 7.633e-3 $\pm$  2.523e-4\\ \hline
$y_{8}$& 521 & 401 & 14 &  \\
& ($0.3 \leq y_8 \leq 0.5$ ) & ($0.4 \leq y_8 \leq 0.55$)  & ($0.5 \leq y_8 \leq 0.65$ ) & 1.009e-3 $\pm$ 4.065e-4\\ \hline
$y_{9}$& 401 & 261 & 12 &  \\
& ($0.01 \leq y_9 \leq 0.03$)  & ($0.025 \leq y_9 \leq 0.035$)  & ($0.03 \leq y_9 \leq 0.05$)  & 3.683e-3 $\pm$ 7.473e-4\\ \hline
$y_{10}$& 451 & 281 & 8 & \\
& ($0.0 \leq y_{10} \leq 0.25$)  & ($0.2 \leq y_{10} \leq 0.35$)  & ($0.3 \leq y_{10} \leq 0.5$)  & 5.388e-3 $\pm$ 4.331e-4\\ \hline
$y_{11}$& 501 & 421 & 12 &  \\
& ($0.0 \leq y_{11} \leq 0.4$  & ($0.3 \leq y_{11} \leq 0.45$)  & ($0.4 \leq y_{11} \leq 0.8$ ) & 3.638e-2 $\pm$ 6.999e-4 \\ \hline
 $y_{12}$& 301 & 201 & 12 & \\
& ($0.0 \leq y_{12} \leq 0.003$  & ($0.002 \leq y_{12} \leq 0.004$ ) & ($0.003 \leq y_{12} \leq 0.006$)  & 6.673e-3  $\pm$ 4.425e-4 \\ 
 \hline
 $y_{13}$& 601 & 501 & 20 &  \\
& ($0.0 \leq y_{13} \leq 0.01$  & ($0.005 \leq y_{13} \leq 0.015$)  & ($0.01 \leq y_{13} \leq 0.03$)  & 1.405e-2 $\pm$ 3.539e-3 \\ \hline
$y_{14}$& 801 & 501 & 20 & \\
& ($0.0 \leq y_{14} \leq 0.3$  & ($0.2 \leq y_{14} \leq 0.35$)  & ($0.3 \leq y_{14} \leq 0.45$)  & 2.993e-3 $\pm$ 5.536e-4\\ \hline
$y_{15}$& 601 & 421 & 14 & \\
& ($0.0 \leq y_{15} \leq 0.25$  & ($0.2 \leq y_{15} \leq 0.3$)  & ($0.25 \leq y_{15} \leq 0.35$)  & 3.699e-3 $\pm$ 6.678e-4 \\ \hline
$y_{16}$& 501 & 301 & 16 &  \\
& ($0.0 \leq y_{16} \leq 0.3$  & ($0.25 \leq y_{16} \leq 0.35$)  & ($0.3 \leq y_{16} \leq 0.5$)  & 9.076e-3 $\pm$ 1.538e-4\\ \hline
$y_{17}$& 301 & 231 & 16 &  \\
& ($0.007 \leq y_{17} \leq 0.01$  & ($0.009 \leq y_{17} \leq 0.014$)  & ($0.012 \leq y_{17} \leq 0.016$)  & 3.584e-2 $\pm$ 7.858e-4\\ \hline
$y_{18}$& 601 & 441 & 20 &  \\
& ($0.0 \leq y_{18} \leq 0.3$  & ($0.24 \leq y_{18} \leq 0.34$ ) & ($0.3 \leq y_{18} \leq 0.4$)  & 5.526e-3 $\pm$ 4.426e-4\\ \hline
$y_{19}$& 801 & 511 & 12 &  \\
& ($0.0 \leq y_{19} \leq 0.2$  & ($0.15 \leq y_{19} \leq 0.3$)  & ($0.25 \leq y_{19} \leq 0.4$)  & 4.411e-3 $\pm$ 5.748e-4\\ \hline
$y_{20}$& 501 & 321 & 18 & \\
& ($0.0 \leq y_{20} \leq 0.35$  & ($0.3 \leq y_{20} \leq 0.4$)  & ($0.35 \leq y_{20} \leq 0.45$)  & 5.308e-3 $\pm$  3.759e-4  \\ \hline
\end{tabular}
\label{table3}
\end{center}
\end{table}

\subsubsection{Example $3$: Syngas (Pure kinetics)}
\label{subsec:example3}

In this example, we consider the kinetics of a skeletal model of syngas for CO/H\textsubscript{$2$} burning. The skeletal mechanism has $n_s=11$ species and $21$ reactions. This fuel is a key subset of higher hydrocarbon fuels, and it represents syngas. The fuel consists of $50$\% CO, $10$\% H\textsubscript{$2$} and $40$\% N\textsubscript{$2$} and the oxidizer consists of $25$\% O\textsubscript{$2$}  and $75$\% N\textsubscript{$2$}. For training, we generated data for different equivalence ratios in the range of $\phi^0 = [0.7, 1.4]$ and different initial temperatures in the range of $T^0= [500, 1250]$ (K). The initial temperature was sampled at 300 uniformly distributed points in this range. The equivalence ratio was sampled at 100 uniformly distributed points in the range shown above. Overall, we performed $300\times 100 = 30\small{,}000$ simulations of the kinetics equation for all the combinations of initial temperature values and equivalence ratios. For time integration, we use explicit fourth-order Runge Kutta with $\Delta t_{chm}= 10^{-8} $ (sec). For this example, to develop an efficient surrogate model, we learn the dynamics with a DeepONet as well as with an autoencoder integrated with DeepONet.\\

\begin{figure}[!ht]
\centering
\includegraphics[width=1\textwidth]{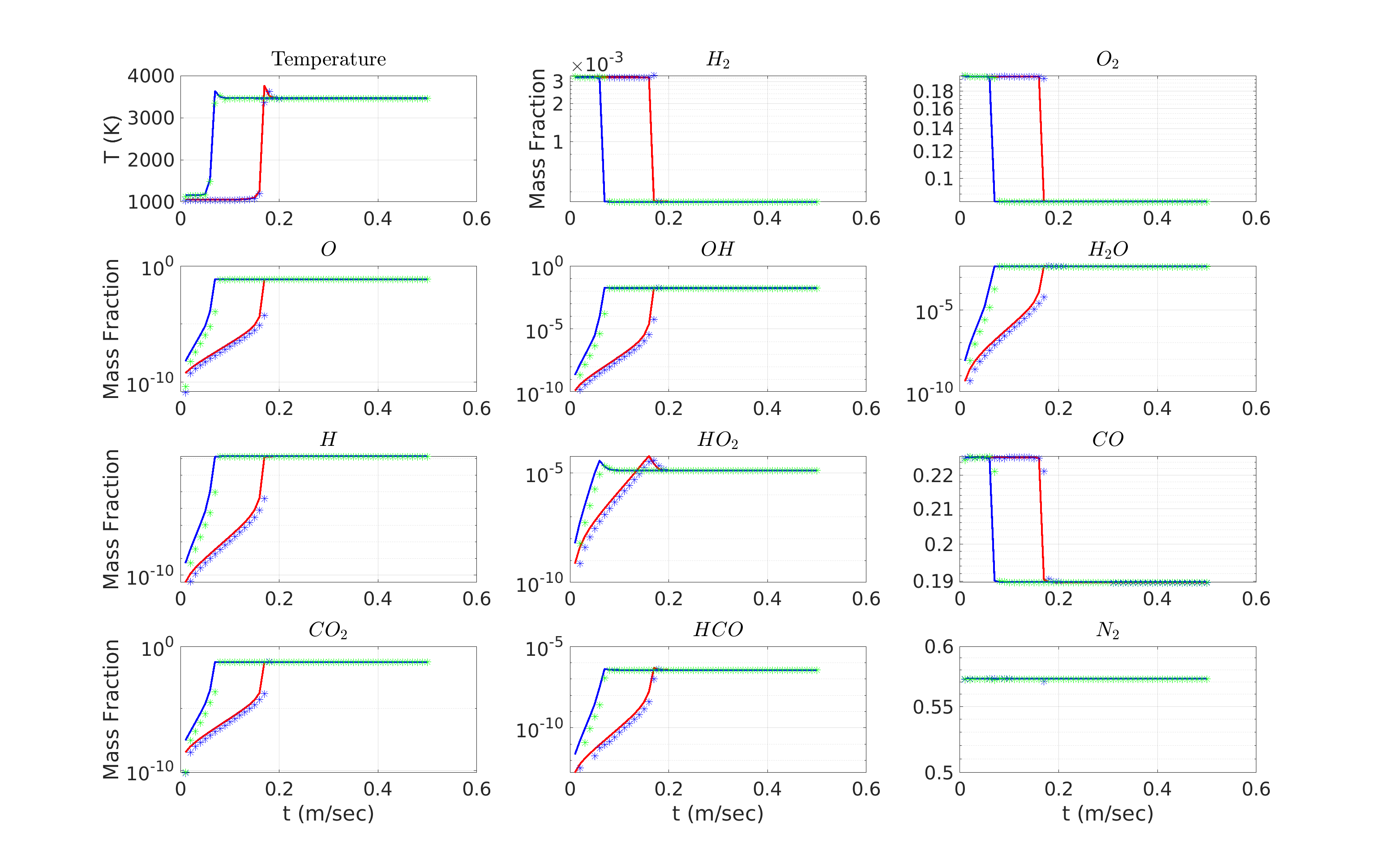}
\caption{DeepONet (without AE) for syngas (pure kinetics): Representative plots for two testing cases with initial temperature, $T_0 = 1051.7241$ (red solid line and blue $*$ markers) and $T_0 = 1155.1724$ (blue solid line and green $*$ markers) and the same equivalence ratio, $\phi_0 = 0.77778$. The solid lines represent the ground truth, and the $*$ markers represent the predicted values for the QoIs.}
\label{fig:prediction1}
\end{figure}

\noindent 
\textbf{Learning the dynamics using DeepONet:} \\
To train the DeepONet, we randomly chose $N_{train} = 10\small{,}000$ realizations (out of $30\small{,}000$) of $\mathbf{x}_i$, such that each realization consists of the mass fractions, $Y_i^j$ of $n_s = 11$ species, where $j = \{1,2, \ldots, 11\}$, temperature, $T_i$ at time step $t_i$, where $i =\{1, 2, \ldots, N_{train}\}$ and the parameterized initial conditions, $\phi_i^0$ and $T_i^0$, which denote the initial equivalence ratio and the initial temperature, respectively. Hence, $\mathbf{x}_i = \{Y_i^1, Y_i^2, \ldots, Y_i^{n_s}, T_i, \phi_i^0, T_i^0\}$ and $\mathbf X_{train} = \{\mathbf x_i\}_{i=1}^{N_{train}}$. The trunk net inputs the temporal coordinate which is the time steps ahead of the initial time step at which the species concentration and the temperature are to be computed. The input and the output space of the training data are normalized using QoI specific mean, $\mu_i$, and standard deviation, $\sigma_i$ of the logarithmic value of the corresponding values. For testing the accuracy of the trained DeepONet model, we sampled $N_{test} = 30\small{,}000$ realizations from the generated data. The DeepONet model trained with Eq.~\ref{eq:output1_deeponets}, reported a mean relative $\mathcal L_2$ error of $0.3\%$ for the unseen test dataset. The plots for the one sample testing case are presented in Fig.~\ref{fig:prediction1}.\\

\begin{figure}[!t]
\centering
\includegraphics[width=\textwidth]{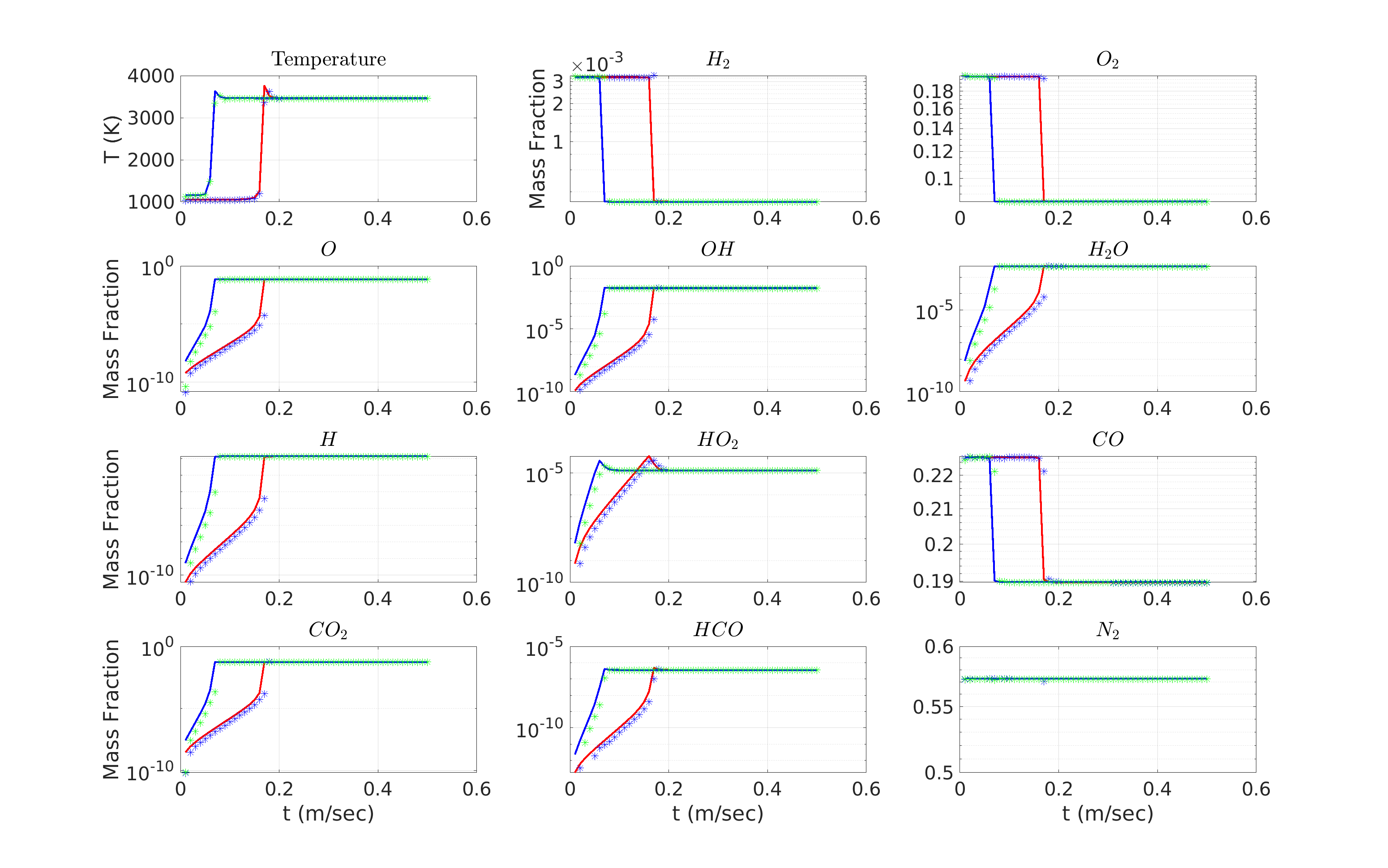}
\caption{AE$+$DeepONet results for syngas (Pure kinetics): Representative plots for two testing cases with $T_0 = 1051.7241$ (red solid line and blue $*$ markers) and $T_0 = 1155.1724$ (blue solid line and green $*$ markers) with the same equivalence ratio, $\phi_0 = 0.77778$. The solid lines represent the ground truth, and the $*$ markers represent the predicted values for the QoIs.}
\label{fig:prediction2}
\end{figure}

\begin{figure}[!ht]
\centering
\includegraphics[width=0.6\textwidth, trim=3cm 8.5cm 4.5cm 8.81cm, clip]{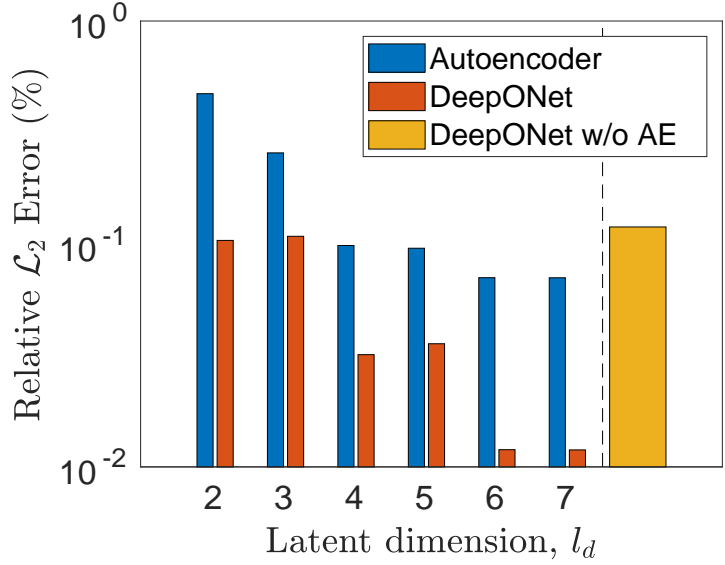}
\caption{Relative $\mathcal L_2$ errors obtained for the test cases while learning the dynamics of syngas (Pure kinetics) with AE$+$DeepONet and DeepONet framework. The computed error for the AE$+$DeepONet framework consists of the reconstruction error of the AE (shown with blue bars) and the generalization error of the DeepONet (shown with red bars), which has been trained on the latent space obtained from the AE. The plot shows the reconstruction error of AE and the generalization error of DeepONet with different latent dimensions, $l_d$. Furthermore, the accuracy of the AE$+$DeepONet framework is compared with the accuracy when DeepONet is trained on the full dimension (shown with a yellow bar).}
\label{fig:example3_ae}
\end{figure}

\noindent
\textbf{Learning the dynamics using Autoencoder $+$ DeepONet:}\\
Now we explore the methodology discussed in Section~\ref{subsec:ae_deeponet} to build an efficient surrogate model for approximating the dynamics of the chemical reaction. For the syngas combustion with an original dimension of $n_s+1 = 11+1$, we examine the dimension of the latent versus different levels of accuracy loss. To train the AE, we sampled $N_{train} = 30\small{,}000$ realizations of $\mathbf{x}_i$, such that each realization consists of the mass fractions, $Y_i^j$ of $n_s = 11$ species, where $j = \{1,2, \ldots, 11\}$ and temperature, $T_i$ at time step $t_i$. The training and testing data is normalized with a mean and standard deviation on the logarithmic value of the values. Fig.~\ref{fig:example3_ae} shows the errors with the reconstruction error of the AE for different $n_z = \{1,2,...,6\}$ which is computed $N_{test} = 20\small{,}000$ unseen samples. We observe that for $n_z > 6$, the reconstruction error saturates. Therefore, we use $n_z = 6$ as the latent dimension for further training in the DeepONet. To prepare training data for DeepONet, we define $\mathbf Z = \{\mathbf z_1, \mathbf x_2, \dots, \mathbf z_{N_{train}}\}$, where $\mathbf z_i = \{z_i^1,z_i^2,...,z_i^{n_z}, \phi_i^0, T_i^0\}$. In this setup, we learn the mapping $\mathcal G_{\boldsymbol \theta}: \mathcal{R}^{n_z+2} \rightarrow \mathcal{R}^{n_t\times n_z}$. We perform a standard scaling before we train the DeepONet with $\mathbf Z$. The DeepONet model trained with Eq.~\ref{eq:output1_deeponets}, reported a mean relative $\mathcal L_2$ error of $0.01\%$ for the unseen test dataset on the latent dimension. The solution predicted by DeepONet on the latent dimension is used as inputs to the pre-trained decoder and the mean relative $\mathcal L_2$ error of the integrated setup is $0.1\%$. The plots for the one testing case are presented in Fig.\ref{fig:prediction2}. Additionally, we report the training cost of two approaches discussed previously for this problem in Table~\ref{table:computationaltime-MLAE} (first row) and show that learning the dynamics using the AE$+$DeepONet framework is significantly cheaper than training the DeepONet on the full-dimension dataset. After training the networks, the inference time for testing new samples is typically very fast, taking only a fraction of a second, $\sim\mathcal{O}(10^{-2})$ second.

\subsubsection{Example $4$: Syngas (Turbulent flame)}
\label{subsec:example4}

For the final demonstration, we consider the same syngas mechanism as in the previous demonstration for a temporally developing planar CO/H\textsubscript{$2$} jet flame. This turbulent flame has been the subject of previous detailed direct numerical simulation (DNS) \cite{Hawkes2007Scalar,ANGB22,yang2013large,punati2011evaluation,Vo2018MMC,yang2017sensitivity,sen2010large, Aitzhan2021arXiv}. In particular, the problem setup considered here is identical to the one used recently in \cite{Aitzhan2021arXiv}. The configuration as considered here is the two-dimensional version of that in \cite{Hawkes2007Scalar}. The flame is rich with strong flame-turbulence interactions, resulting in local extinction followed by re-ignition.  The jet consists of a central fuel stream of width $H=0.72$mm surrounded by counter-flowing oxidizer streams. The fuel stream is comprised of $50\%$ of CO, $10\%$ H\textsubscript{$2$} and $40\%$ N\textsubscript{$2$} by volume, while oxidizer streams contain $75\%$ N\textsubscript{$2$} and $25\%$ O\textsubscript{$2$}. The initial temperature of both streams is $500$K and thermodynamic pressure is set to $1$ atm. The velocity difference between the two streams is $U = 145$m/s. The fuel stream velocity and the oxidizer stream velocity are $U/2$ and $-U/2$, respectively. The initial conditions for the velocity components and mixture fraction are taken directly from center-plane DNS in \cite{Hawkes2007Scalar}, and then the spatial fields of species and temperature are reconstructed from a flamelet table generated with $\chi = 0.75 \chi_{\text{crit}}$, where $\chi$ and $\chi_{\text{crit}}$ are scalar dissipation rate and its critical value respectively. The boundary conditions are periodic in stream-wise ($x$) and cross-stream-wise ($y$) directions. The Reynolds number based on $U$ and $H$ is $Re=2510$. The sound speeds in the fuel and the oxidizer streams are denoted by $C_1$ and $C_2$, respectively, and the Mach number $Ma = U / \left(C_1 + C_2 \right) \approx 0.16$. The combustion chemistry is modeled via the syngas mechanism \cite{punati2011evaluation} containing $11$ species with $21$ reaction steps.

The computational domain is a rectangle with a length of $L_x = 8.625 $ mm and a height of $L_y = 10.065$ mm, where $ 0 \leq x \leq L_x$ and $ -L_y/2 \leq y \leq L_y/2$.  The DNS of the base reactive flow is conducted via a Fourier spectral solver with $N_x = 576$ and $N_y = 672$ Fourier modes in $x$ and $y$ directions, respectively. Therefore, the total number of grid points is roughly equal to 387,000. This amounts to a uniform mesh in both $x$ and $y$ directions with an approximate size of $\Delta x = \Delta y = 0.015$ mm. Simulations are conducted for the duration $0 \le t \le 26t_j$,  $t_j=H/U$.  The compressible Navier Stokes equations and the species transport equation are solved using the fourth-order Runge Kutta scheme with the constant $\Delta t = 10^{-8}$ (sec). This amounts to roughly 13,000 time steps for the entire simulation. This $\Delta t$ is small enough to allow an explicit time to be used for the integration of the kinetic terms.

Only a subset of the data generated by the DNS is used for DeepONet training. In particular, the training data are taken from a rectangle with the bounds of $0\leq x \leq 8.625$ mm and $-2 \leq y \leq  2$ mm. This selected region includes the entire width of the computational domain (in the $x$ direction) and only a segment of the computational domain in the $y$ direction, which is large enough to encompass the jet stream evolution during the time of the simulation.  Moreover, not all of the DNS grid points within this region are used for DeepONet training. In particular, only $5,000$ grid points that are uniformly distributed in this region are chosen for training. Note that there are roughly $154,000$ DNS grid points within this region. Therefore, the selected points are about 3.25 \% of the DNS grid points in this rectangle.  The DeepONet is trained for $\zeta = \{t_{i}+\hat{t}_1, t_{i}+\hat{t}_2, \dots, t_{i}+\hat{t}_{n_t}\}$, where $n_t=4$ and $t_1=250 \Delta t$, $t_2=500 \Delta t$, $t_3=500 \Delta t$, and $t_4=1,000 \Delta t$.  For preparing the training dataset, we take input to the branch network as the chemical description at the time step following the reset of the CFD solver, to learn the kinetics description after $250,\;500,\;750$, and $1000$ time steps. The details of preparing the training dataset are discussed in subsequent sections.
For testing the trained model, we use a new simulation, where the trained operator network is employed to predict the kinetics description from time $=[0, 13000]\Delta t$, when the kinetics solver is not reset. The output of the DeepONet at the $1000$-th time step is considered an initial condition for the next $1000$ time steps and is taken as an input to the branch network. Details regarding the training and testing of the two frameworks are presented below. \\

\begin{figure} [htpb] 
\centering 
\includegraphics[width=\textwidth]{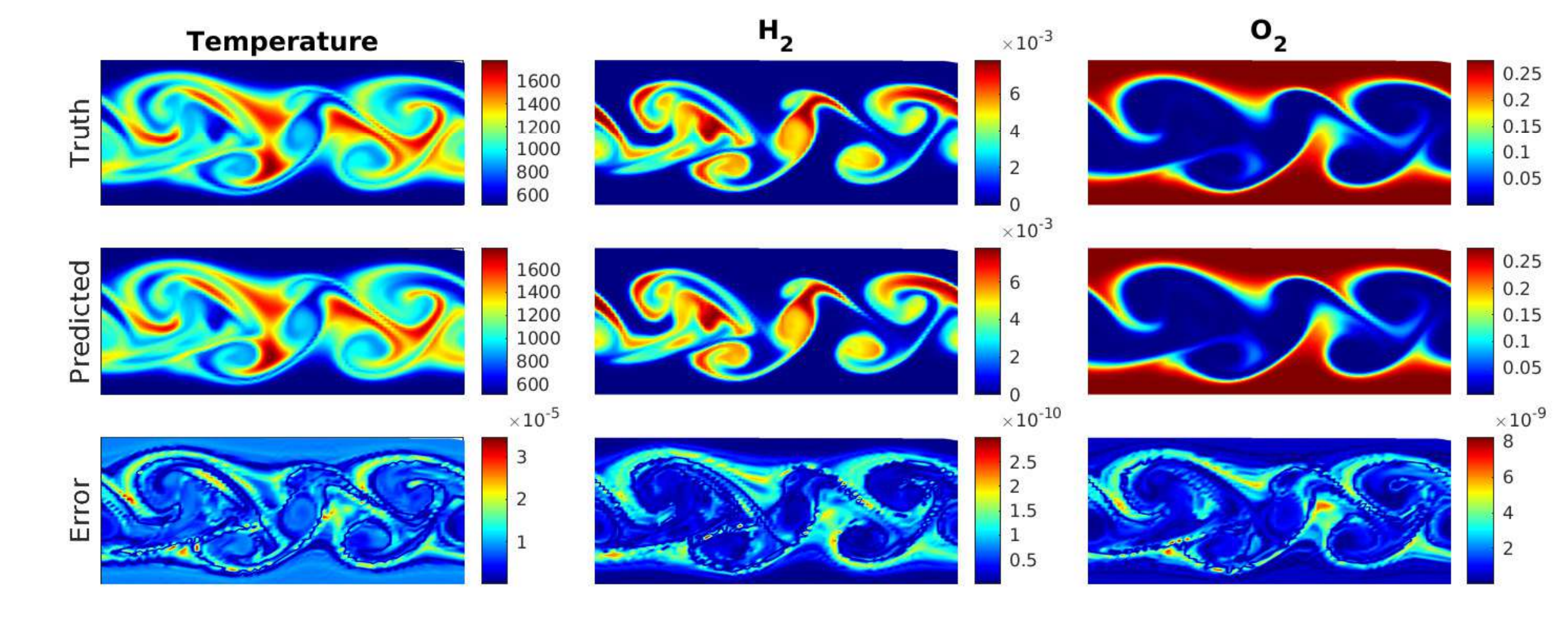}
\includegraphics[width=\textwidth]{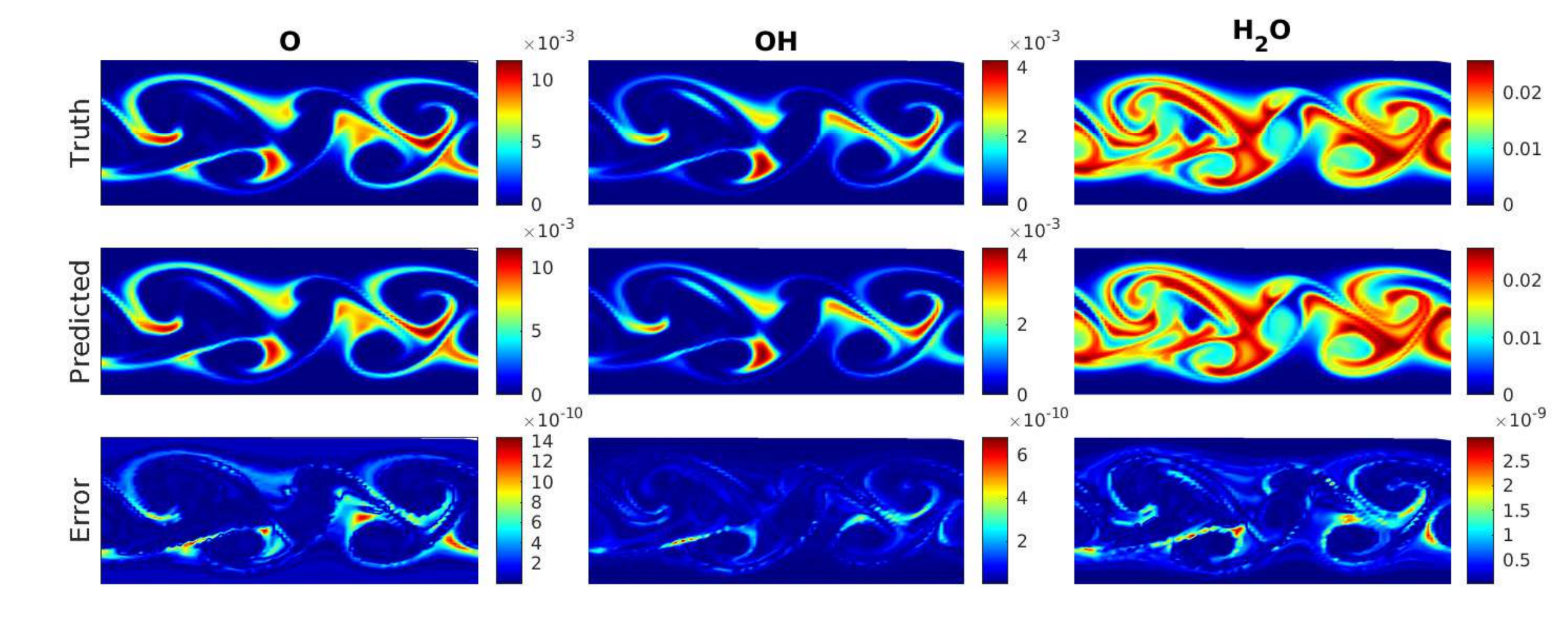}
\caption{DeepONet (without AE) for syngas (turbulent flame): Representative plots of temperature and $5$ species to show the advancement of $1000$ time steps for a given initial condition. The remaining species are shown in Fig.~\ref{fig:prediction3_2}. The plots presented here are for a non-recursive update of the input to the branch network of DeepONet.}
\label{fig:prediction3_1}
\end{figure}

\begin{figure} [htpb] 
\centering 
\includegraphics[width=\textwidth]{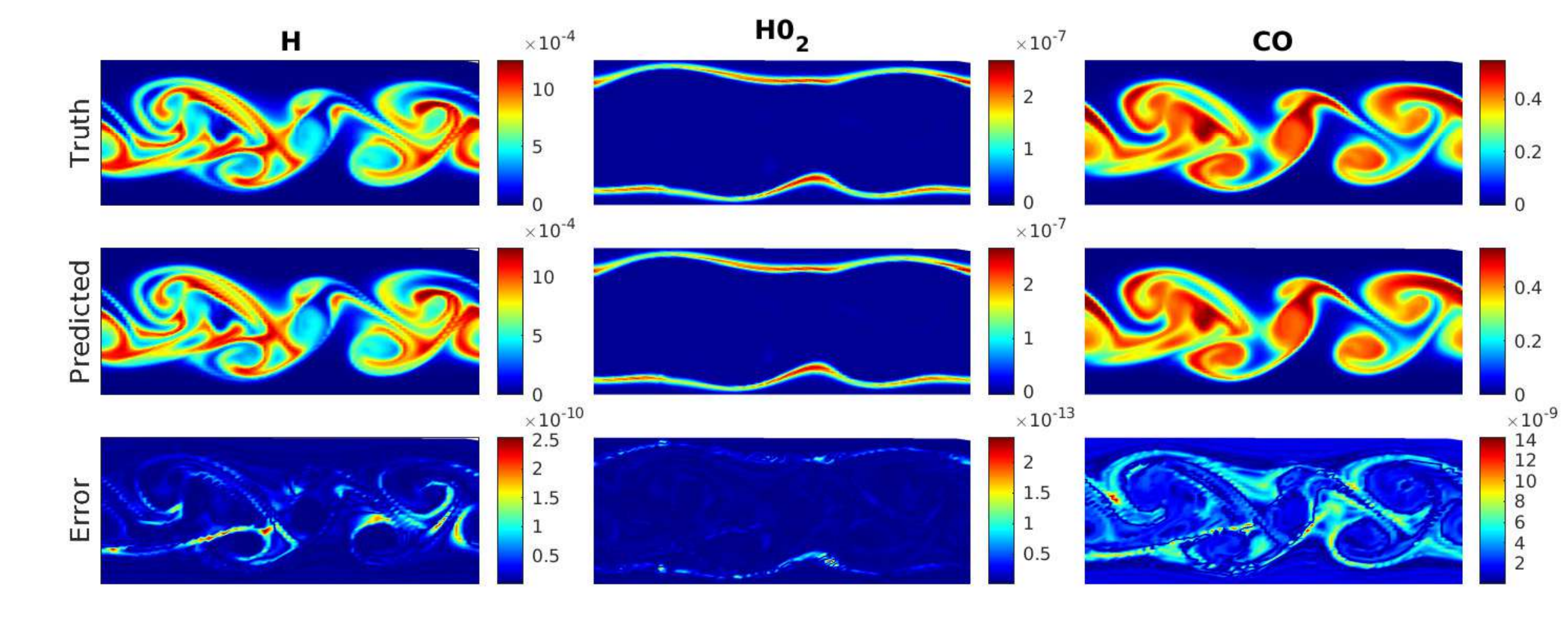}
\includegraphics[trim = 0cm 0cm 8cm 0cm, clip, width=0.68\textwidth]{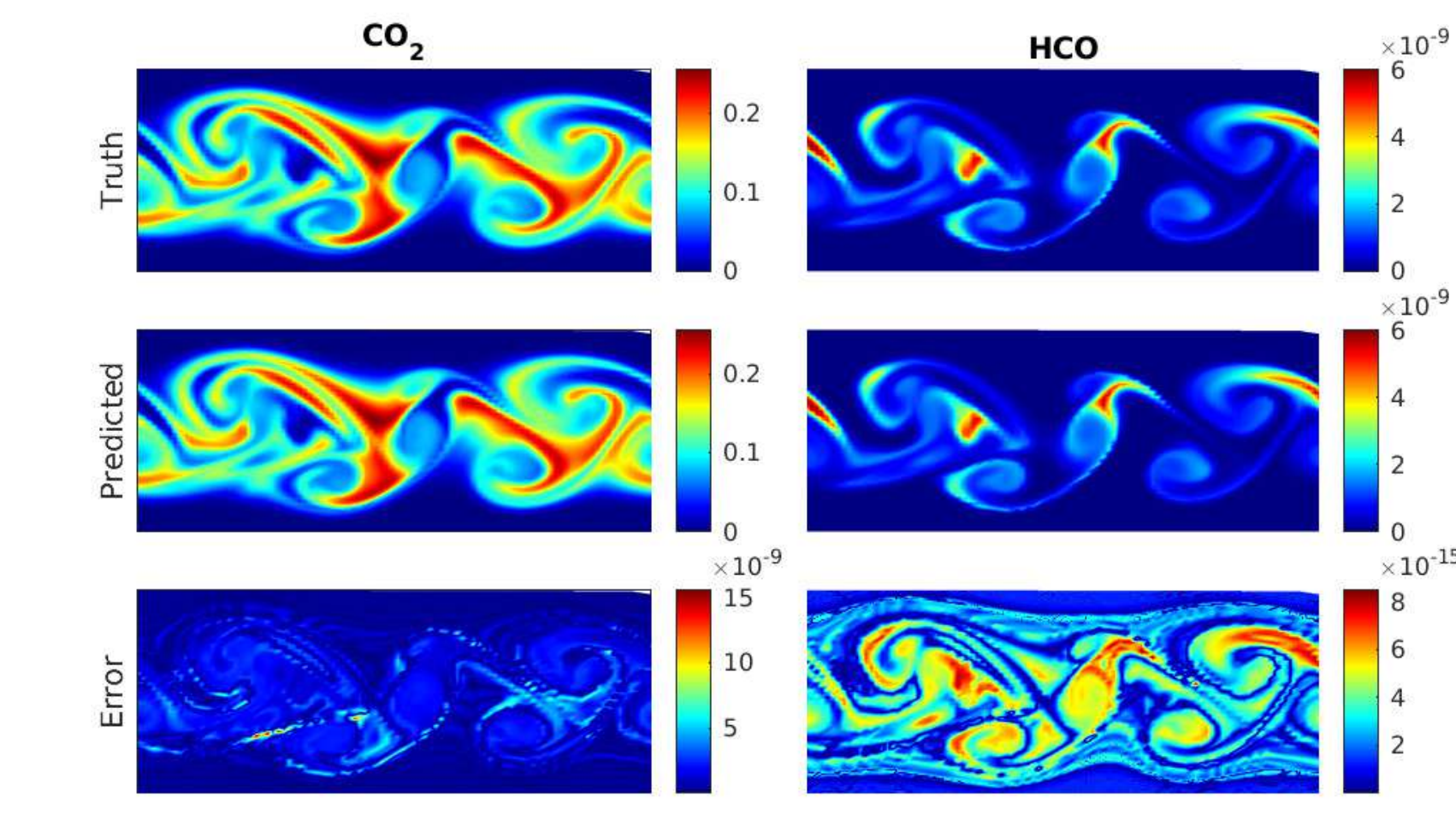}
\caption{DeepONet (without AE) for syngas (turbulent flame): Representative plots of $5$ species to show the advancement of $1000$ timesteps for a given initial condition. The species shown in this plot are in addition to the species presented in Fig.~\ref{fig:prediction3_1}. The plots presented here are for a non-recursive update of the input to the branch network of DeepONet.}
\label{fig:prediction3_2}
\end{figure}

\begin{figure} [htpb] 
\centering 
\includegraphics[trim=4cm 8.4cm 4cm 9cm,clip,scale=0.67]{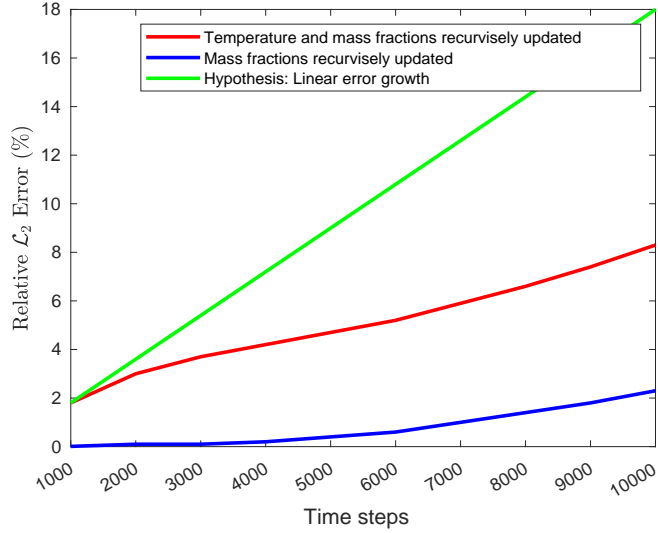}
\caption{Error growth in DeepONet (without AE) for syngas (turbulent flame): The plot presents the growth of error when the inputs of the model are recursively updated with the outputs of the model from the previous timestep. In this plot, the red line depicts a scenario when the temperature, as well as the mass fractions, are recursively updated from outputs of the model, while the blue line depicts a scenario when only the mass fractions of the species are updated but the temperature of the current step is provided from the ground truth. The green line is shown as a benchmark line to show that in both scenarios, the error growth is sub-linear.}
\label{fig:err_accumulation}
\end{figure}

\noindent
\textbf{Learning the dynamics using DeepONet:} 
\noindent To train the DeepONet, we sample $N_{train} = N_t\times N_p$ realizations, where $N_t =13$ is the number of times the CFD solver has been reset and $N_p = 5000$ is the discretization of the domain of study. Each realization $\mathbf{x}_i \in \mathbf{X}$ comprises of the mass fractions, $Y_i^j$ of $n_s = 11$ species, where $j = \{1,2, \ldots, 11\}$ and temperature, $T_i$, where $i =\{1, 2, \ldots, N_{train}\}$. Hence, $\mathbf{x}_i = \{Y_i^1, Y_i^2, \ldots, Y_i^{n_s}, T_i\}$ and $\mathbf X_{train} = \{\mathbf x_i\}_{i=1}^{N_{train}}$. The trunk net inputs the temporal coordinates, which are the time steps ahead of the initial time step at which the species concentration and the temperature are to be computed. The input and the output space of the training data are normalized using QoI specific mean, $\mu_i$, and standard deviation, $\sigma_i$ of the corresponding species. For testing the accuracy of the trained DeepONet model, we sampled $N_{test} = N_{t,test}\times N_{p}$ realizations, where $N_{t,test} = 50\small{,}000$ time steps. The testing is carried out as two experiments. In the first experiment, we consider $N_{test}$ initial conditions as input to the branch net and predict the dynamics of the setup which would be $250,\;500,\;750$, and $1000$ time steps ahead. This experiment reported a mean relative $\mathcal L_2$ error of $0.03\%$ for the test dataset. The plots for the one sample testing case is presented in Figs.~\ref{fig:prediction3_1} and \ref{fig:prediction3_2}. In the second experiment, we recursively obtain the kinetics description of t $=\{2, 50000\}$, by starting with the first time step as inputs to the branch net and then using the prediction of the trained model as input to the DeepONet to predict the next time steps. In this experiment, we observe the accumulation of errors as we march ahead in time. In Fig.~\ref{fig:err_accumulation}, we present the error accumulation plot for two different test cases. The error growth curve shown with the red line is a scenario when the input to the DeepONet model is recursively updated with the temperature and the mass fractions obtained as outputs from the DeepONet at an earlier time step. Additionally, the blue line denotes the scenario for obtaining the predictions at $1000$ timesteps ahead, only the mass fractions of the species are recursively updated as inputs to the model, but the ground truth of temperature is provided from the labeled dataset for the current timestep. The results indicate that some sort of information needs to be updated in the model to correct the dynamics and approximate the time evolution accurately. To control the error growth over time, one approach is to employ the hybrid scheme proposed in \cite{oommen2022learning}. The approach involves iterating between the high-fidelity solver and the surrogate model, such that the high-fidelity model acts as a corrector of the dynamics and the surrogate model helps the leap in time. Furthermore, in Fig.~\ref{fig:err_accumulation}, we also show that the error growth in both test cases is sub-linear. Additionally, exploring the ideas of operator-level transfer learning \cite{zhu2022reliable,goswami2022deep} can also help in keeping the error growth within a certain pre-decided bound. This will be considered in future work. 

\bigbreak
\noindent
\textbf{Learning the dynamics using Autoencoder $+$ DeepONet:} 

\begin{figure} [htpb] 
\centering 
\includegraphics[width=\textwidth]{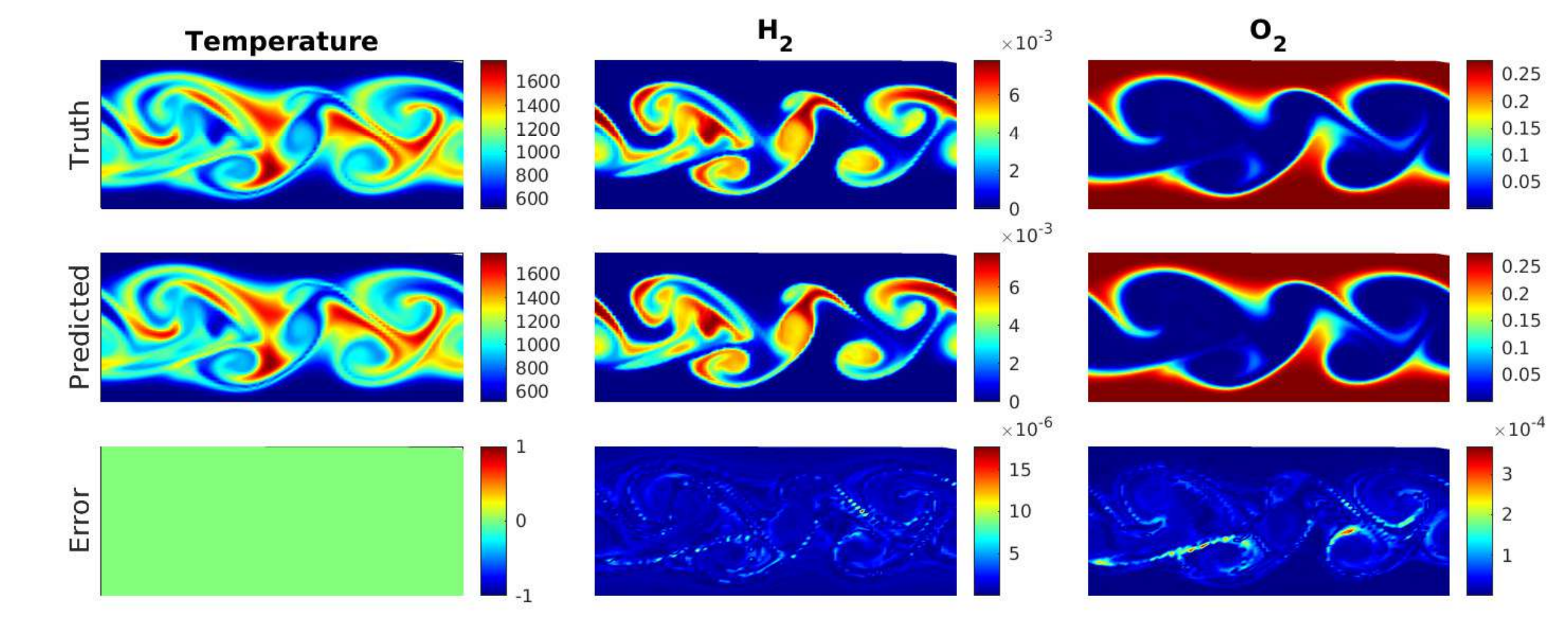}
\includegraphics[width=\textwidth]{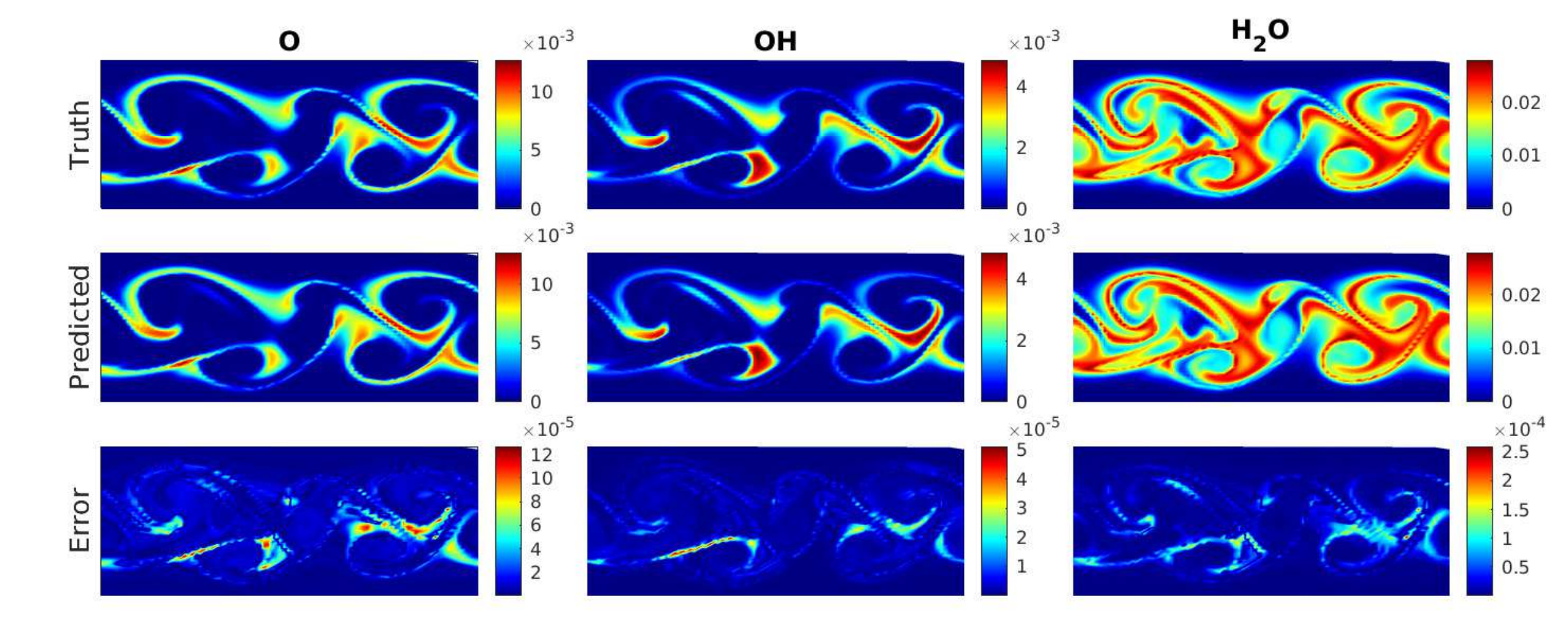}
\caption{AE $+$ DeepONet for syngas (turbulent flame): Representative plots of temperature and $5$ species to show the advancement of $1000$ time steps for a given initial condition. The remaining species are shown in Fig.~\ref{fig:prediction4_2}. The plots presented here correspond to a non-recursive update of the input to the branch network of DeepONet.}
\label{fig:prediction4_1}
\end{figure}

\begin{figure} [htpb] 
\centering 
\includegraphics[width=\textwidth]{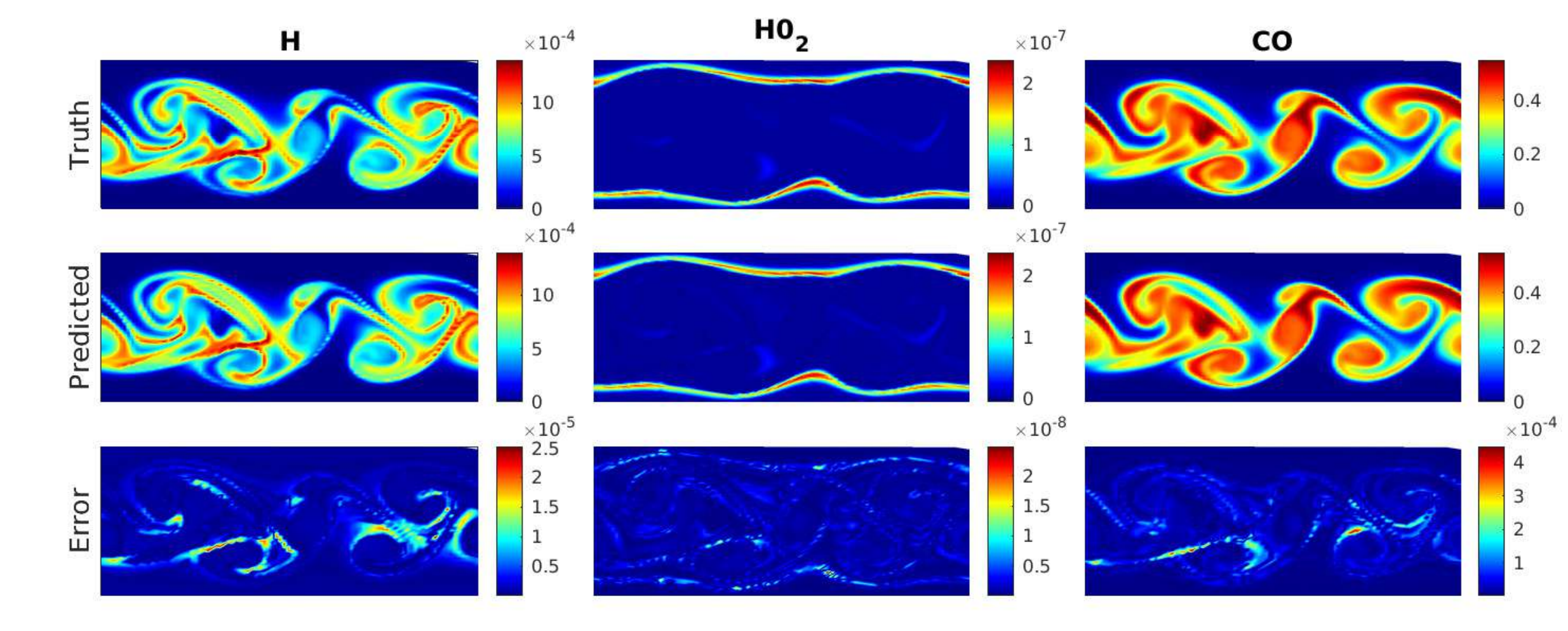}
\includegraphics[trim = 0cm 0cm 8cm 0cm, clip, width=0.68\textwidth]{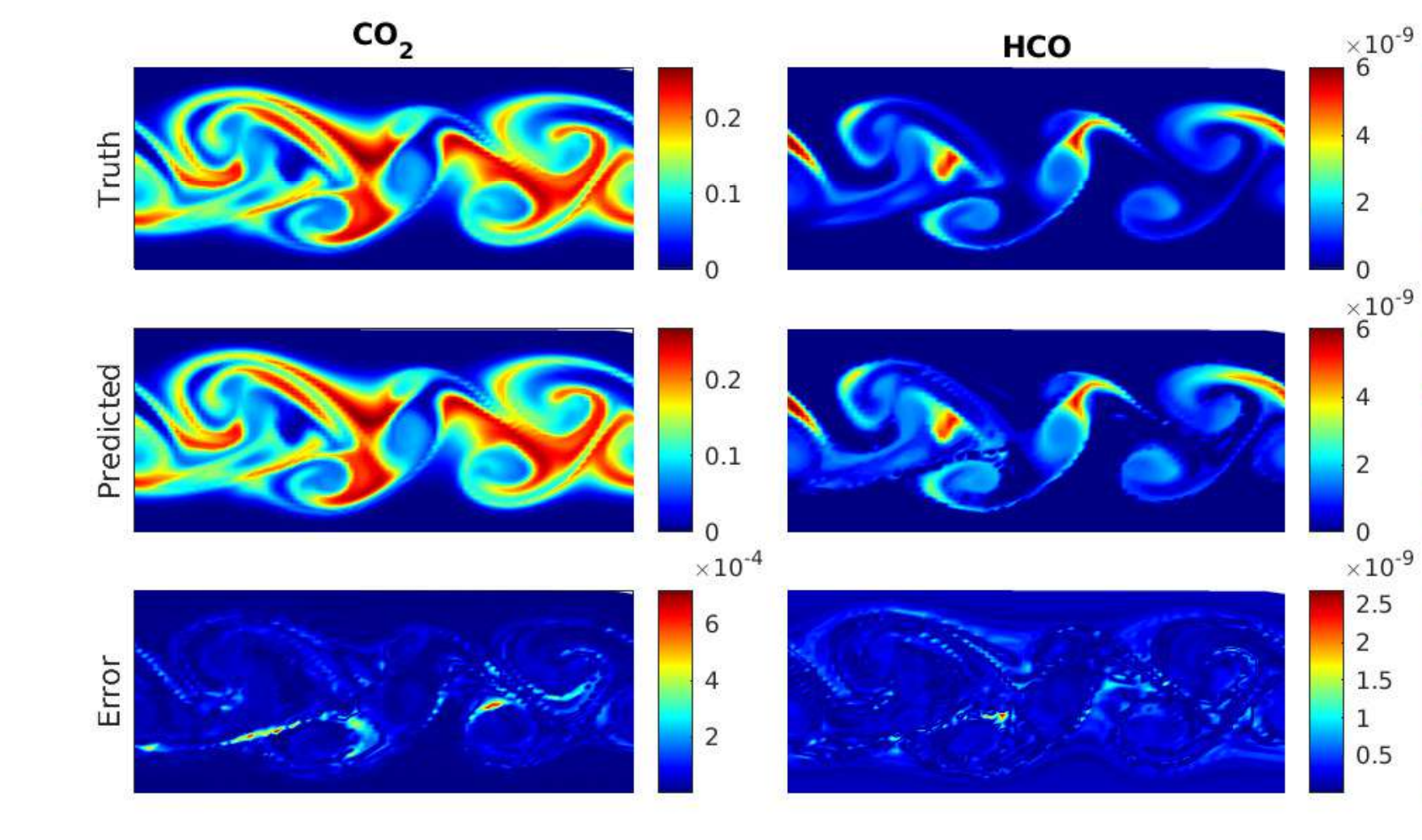}
\caption{AE $+$ DeepONet for syngas (turbulent flame): Representative plots of $5$ species to show the advancement of $1000$ timesteps for a given initial condition. The species shown in this plot are in addition to the species presented in Fig.~\ref{fig:prediction4_1}. The plots presented here correspond to a non-recursive update of the input to the branch network of DeepONet.} 
\label{fig:prediction4_2}
\end{figure}

\noindent Now we explore the framework discussed in Section~\ref{subsec:ae_deeponet} to build an efficient and robust surrogate model for approximating the dynamics of the chemical reaction in latent space. For this example, the original dimensionality is $n_s+1 = 11+1$. To train the AE, we used $0.8\times N_{train}$ realizations of $\mathbf{X}$ and tested the reconstruction error on $0.2\times N_{train}$ realizations. The training and testing data is normalized with a mean and standard deviation on the logarithmic value of the training data. In this example, we observed that for $n_z > 2$, the reconstruction error saturates. So we use $n_z = 2$ as the latent dimension for further training in the DeepONet. To prepare training data for DeepONet, we define $\mathbf Z = \{\mathbf z_1, \mathbf x_2, \dots, \mathbf z_{N_{train}}\}$, where $\mathbf z_i = \{z_i^1,z_i^2\}$, which is the latent representation of the training data discussed above. Essentially, we learn the mapping $\mathcal G_{\boldsymbol \theta}: \mathcal{R}^{n_z} \rightarrow \mathcal{R}^{n_t\times n_z}$. We perform a standard scaling on $\mathbf Z$ before we train the DeepONet. The DeepONet model trained with Eq.~\ref{eq:output1_deeponets}, reported a mean relative $\mathcal L_2$ error of $0.01\%$ for the test dataset on the latent dimension. The solution predicted by DeepONet on the latent dimension is used as inputs to the pre-trained decoder and the mean relative $\mathcal L_2$ error of the integrated surrogate model is $0.01\%$. The plots for the one sample testing case are  presented in Figs.~\ref{fig:prediction4_1} and \ref{fig:prediction4_2}. Finally, the training cost of the two approaches used for this problem is shown in Table~\ref{table:computationaltime-MLAE} (second row). Consistent with the previous example, learning the dynamics using the AE$+$DeepONet framework is significantly cheaper than training the DeepONet on the full-dimension dataset. Once the networks are trained, the inference time on unknown test samples is $\sim\mathcal{O}(10^{-2})$ seconds.

\begin{table}[ht!]
\begin{center}
\caption{Comparison of the computational training time in seconds (s) for training DeepONet and AE$+$DeepONet frameworks across Examples $3$ and $4$ on an NVIDIA $A6000$ GPU. Once the networks are trained, the inference time on unknown test samples is of the order of $10^{-2}$ seconds.}
\label{table:comp_time}
\begin{tabular}{ l c c } 
\hline
Test Case & DeepONet & AE$+$DeepONet \\
\hline
Example $3$ & $33\small{,}443$ s & $5\small{,}915$ s \\ 
Example $4$ & $13\small{,}506$ s & $3\small{,}566$ s \\  
\hline
\end{tabular}
\label{table:computationaltime-MLAE}
\end{center}
\end{table}

\section{Summary}
In this study, we utilized the Deep Operator Network (DeepONet) to solve stiff problems, with a focus on challenging chemical kinetics equations. The primary objective is to discover a solution propagator for time advancement that could exceed chemistry time scales by several orders of magnitude. We accomplished this by employing DeepONet, as well as its extensions, including the newly proposed Partition-of-Unity-based DeepONet and an autoencoder-integrated DeepONet. The neural operator-based approach has several advantages, with one of the most significant being the ability to train DeepONet offline, and later this model can be employed to approximate the solution for any arbitrary time advancements. In the first two examples, we solved the ROBERS problem, which includes three species, and the POLLU problem with $20$ species and $25$ reactions, and the operator was examined for its ability to extrapolate in the out-of-distribution zone. The results obtained show the accuracy of the DeepONet, which can be efficiently used as a solution propagator. In the third and the fourth example, we investigated the pure kinetics and the temporally developing turbulent flame of a skeletal model of syngas for $CO/H_2$ burning, which involved $11$ species and $21$ reactions. The results of the DeepONet and autoencoder-based DeepONet frameworks show high accuracy. Moreover, compared to traditional CFD solvers for stiff-chemical kinetics, DeepONet is computationally very efficient. In future work, our aim is to integrate the DeepONet framework into the actual CFD code for solving such stiff chemically reacting problems, which can drastically reduce the computational cost. 

\subsection*{\textbf{Data and materials availability}}
All the associated codes accompanying this manuscript will be made publicly available upon the acceptance of the manuscript.

\subsection*{\textbf{Acknowledgement}}
We would like to acknowledge the financial support of Small Business Technology Transfer (STTR) program, USA (Grant No: GR5291245).

\bibliographystyle{elsarticle-num}
\bibliography{ref}
\end{document}